\documentclass[journal=cmatex,manuscript=article,layout=traditional]{achemso}

\usepackage{graphicx}
\usepackage{dcolumn}
\usepackage{bm}
\usepackage{hyperref}
\usepackage{multirow}

\usepackage{xcolor} 

	\title{\normalsize Incommensurate structural and magnetic modulations in potassium-rich cryptomelane, K$_x$Mn$_8$O$_{16}$ ($x\approx1.45$)}

\author{
	Liam A. V. Nagle-Cocco
}
\affiliation{Cavendish Laboratory, University of Cambridge, Cambridge, CB3 0US, United Kingdom.}
\alsoaffiliation{Yusuf Hamied Department of Chemistry, University of Cambridge, Cambridge, CB2 1EW, United Kingdom.}
\alsoaffiliation{Stanford Synchrotron Radiation Lightsource, SLAC National Accelerator Laboratory, Menlo Park, California 94025, United States.}
\email{lnc@stanford.edu}

\author{
	Joshua D. Bocarsly
	\footnote{Present address: Department of Chemistry \textit{and} Texas Center for Superconductivity, University of Houston, Houston, Texas 77004, United States of America.}~
}
\affiliation{Cavendish Laboratory, University of Cambridge, Cambridge, CB3 0US, United Kingdom.}
\alsoaffiliation{Yusuf Hamied Department of Chemistry, University of Cambridge, Cambridge, CB2 1EW, United Kingdom.}

\author{
	Krishnakanth Sada
	\footnote{Present address: Xerion Advanced Battery Corporation, Kettering, Ohio 45420, United States of America.}~
}
\affiliation{Faraday Materials Laboratory (FaMaL), Materials Research Centre, Indian Institute of Science, C. V. Raman Avenue, Bangalore, 560012, India.}
\alsoaffiliation{Cavendish Laboratory, University of Cambridge, Cambridge, CB3 0US, United Kingdom.}

\author{Nicola D. Kelly}
\affiliation{Cavendish Laboratory, University of Cambridge, Cambridge, CB3 0US, United Kingdom.}

\author{Mathias A. Kiefer}
\affiliation{Stanford Synchrotron Radiation Lightsource, SLAC National Accelerator Laboratory, Menlo Park, California 94025, United States.}

\author{Emmanuelle Suard}
\affiliation{Institut Laue-Langevin, 71 avenue des Martyrs, CS 20156, cedex 9, Grenoble 38042, France.}

\author{Sarah J. Day}
\affiliation{Diamond Light Source, Ltd., Harwell Science and Innovation Campus, Didcot, Oxfordshire OX11 0DE, United Kingdom.}

\author{Cheng Liu}
\affiliation{Cavendish Laboratory, University of Cambridge, Cambridge, CB3 0US, United Kingdom.}

\author{Clare P. Grey}
\affiliation{Yusuf Hamied Department of Chemistry, University of Cambridge, Cambridge, CB2 1EW, United Kingdom.}

\author{Prabeer Barpanda}
\affiliation{Faraday Materials Laboratory (FaMaL), Materials Research Centre, Indian Institute of Science, C. V. Raman Avenue, Bangalore, 560012, India.}

\author{Clemens Ritter}
\affiliation{Institut Laue-Langevin, 71 avenue des Martyrs, CS 20156, cedex 9, Grenoble 38042, France.}

\author{
	Si\^an E. Dutton
}
\affiliation{Cavendish Laboratory, University of Cambridge, Cambridge, CB3 0US, United Kingdom.}
\email{sed33@cam.ac.uk}

\newcommand*{\citen}[1]{%
	\begingroup
	\romannumeral-`\x 
	\setcitestyle{numbers}%
	\cite{#1}%
	\romannumeral-`\x 
	\setcitestyle{super}%
	\endgroup
}

\begin{document}

\begin{tocentry}

    \includegraphics[scale=0.1]{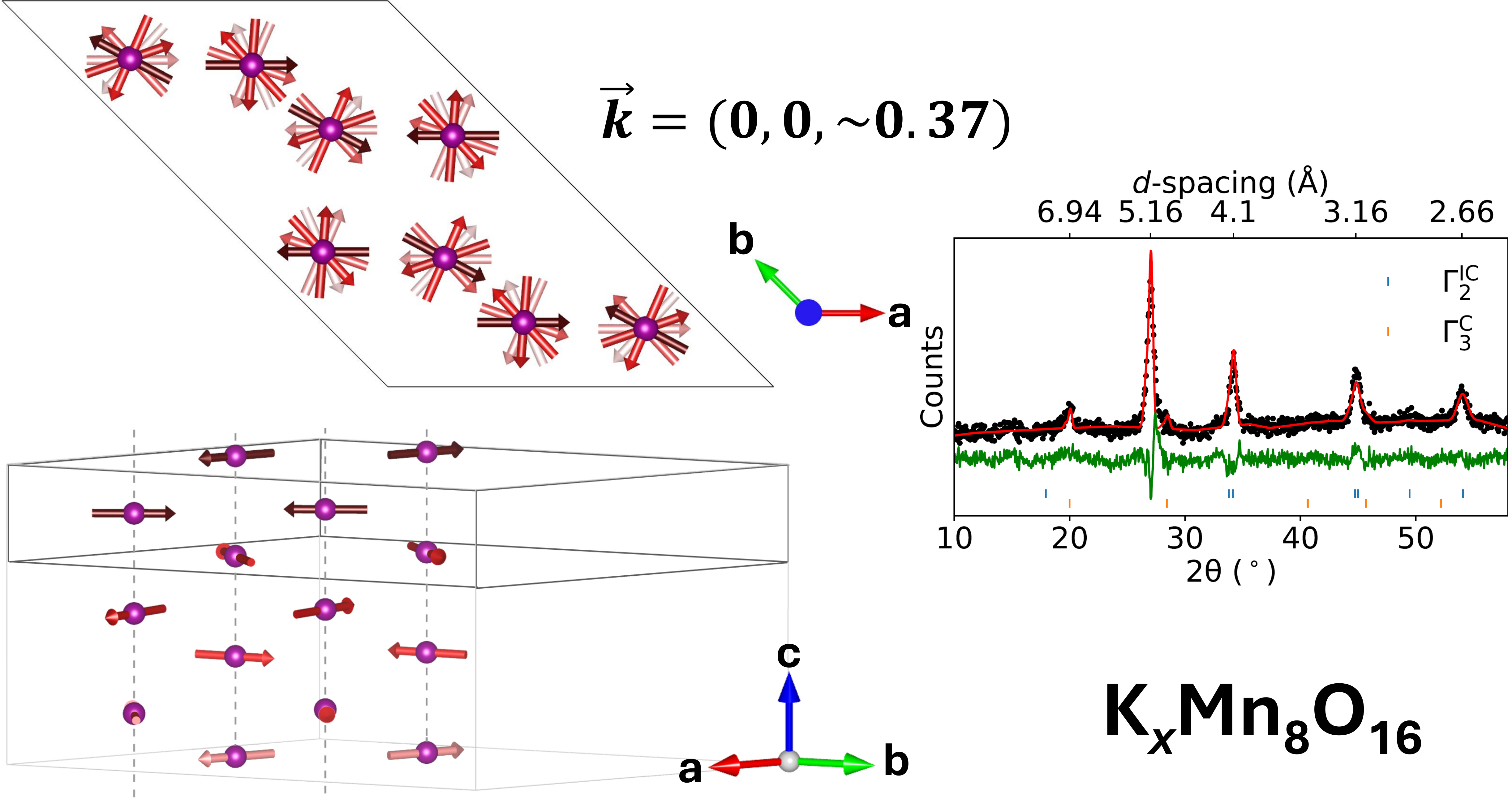}

\end{tocentry}

	\clearpage
	\begin{abstract}
	Cryptomelane is a hollandite-like material consisting of K$^+$ cations in an $\alpha$-MnO$_2$ tunnel-like crystallographic motif. 
	Cryptomelane with stoichiometry K$_x$Mn$_8$O$_{16}$ ($x\approx1.45$) has been synthesized and its magnetic properties investigated using variable-temperature magnetic susceptibility, heat capacity, and neutron powder diffraction. 
	Three distinct transitions at $T_1=184$\,K, $T_2=54.5$\,K, and $T_3=24$\,K are observed. 
	At $T_1$ there is a subtle tetragonal$\rightarrow$monoclinic transition associated with emergence of a set of non-magnetic superstructure peaks indexable to a $\vec{k}_\mathrm{struc}\approx0.74\vec{c^*}$ incommensurate modulation  parallel to the $\alpha$-MnO$_2$ tunnels. 
    Our findings are consistent with a relation previously reported in titanate hollandites, that $x\approx2|\vec{k}_\mathrm{struc}|$. 
	Magnetic Bragg peaks emerge below $T_2=54.5$\,K, and their positions indicate an incommensurate modulated magnetic structure. 
	The model consistent with the data is a dual-$\vec{k}_\mathrm{mag}$ structure with a ferromagnetic $|\vec{k}_\mathrm{mag}|=0$ component and an incommensurate $\vec{k}_\mathrm{mag}\approx0.37\vec{c^*}$, with the latter most likely to be helical. 
	The period of oscillation of the incommensurate magnetic component is in line with predictions based on a Heisenberg spin Hamiltonian [Mandal \textit{et al}. Phys. Rev. B 90, 104420 (2014)]. 
	Below $T_3=24$\,K, there is a magnetic transition, which gives rise to a different set of magnetic Bragg peaks indicative of a highly complex magnetic structure.
	\end{abstract}

	\maketitle

	\section{\label{sec:level1}Introduction}

\begin{figure*}[t]
		\includegraphics{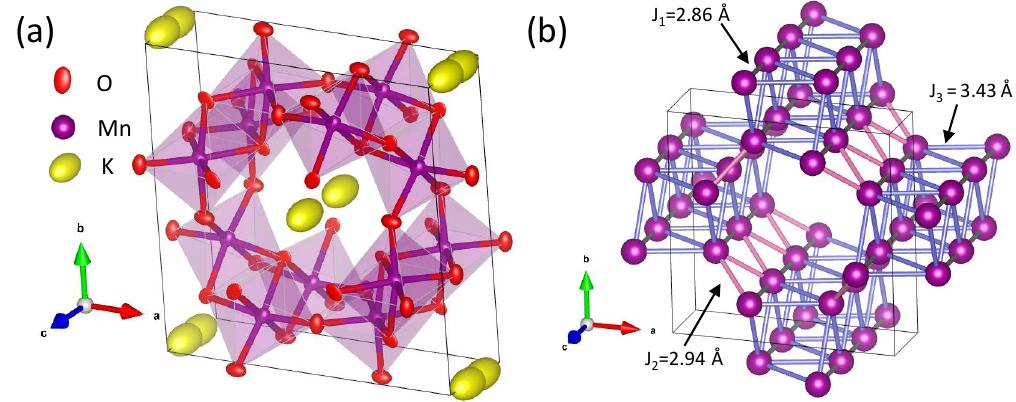}
		\caption{
			\label{fig_struc}
			(a) Crystallographic structure of cryptomelane in the $I4/m$ space group, as viewed at a slight angle from the $c$-axis, with atoms displayed as ellipsoids of 95\% probability. 
			Ellipsoids are constructed based on refined anisotropic atomic displacement parameters for KMO-S1. 
			Red: O; purple: Mn; yellow: K. 
			(b) The Mn lattice of cryptomelane only, showing the three nearest Mn-Mn interactions. 
			Gray: $J_1$, pink: $J_2$, purple: $J_3$. 
			The exchange interactions follow the notation of Crespo \textit{et al.}~\cite{crespo2013competing}
		}
	\end{figure*}

	$\alpha$-MnO$_2$ is a polymorph of Mn$^{4+}$O$_2$ comprised of corner-sharing rutile-like ribbons of MnO$_6$ octahedra which share edges to form a tunnel structure~\cite{bystrom1950crystal}. 
	This structural motif has led to significant interest because in addition to doping within the MnO$_2$ framework~\cite{liu2022tuning}, it is also possible to incorporate dopants into the tunnels. 
	$\alpha$-MnO$_2$ is commonly studied in an $X$-doped form, $X_x$MnO$_{2}$ with $X$ a large cation like K$^+$, Ba$^{2+}$, or Na$^+$~\cite{richmond1942cryptomelane,bystrom1950crystal,lan2011synthesis}. 
	In this work we focus on the K-doped variant, K$_x$Mn$^{3+}_x$Mn$^{4+}_{8-x}$O$_{16}$, which is a naturally-occurring mineral called cryptomelane, and contains mixed Mn valence. 
	K$^+$ cations can be incorporated into the tunnel structure to a maximum of $x=2$, as shown in Figure~\ref{fig_struc}(a), with the doping limited by the occupancy of K in the tunnels. 
	Due to the large tunnels ($\sim$6\,\AA{} in diameter), cryptomelane has significant potential for various applications. 
	For example, it can serve as the positive electrode in K-ion~\cite{poyraz2017synthesis,chong2018cryptomelane} and Zn-ion~\cite{sada2019cryptomelane} batteries, for water splitting via the oxygen evolution reaction~\cite{antoni2019enhancing}, and for oxidizing organic pollutants such as toluene and ethyl acetate~\cite{luo2008adsorptive,santos2010stability,santos2011mixture,davo2016cuo}. 

	Cryptomelane has been studied extensively for its magnetic properties~\cite{strobel1984thermal,suib1994magnetic,sato1997magnetism,sato1999charge,luo2009spin,luo2010tuning,crespo2013competing,mandal2014incommensurate,tseng2015magnetic,barudvzija2016structural,barudvzija2020magnetic}, which stem from its frustration-prone structure. 
	The frustration arises due to competition between the three neighboring Mn-Mn magnetic interactions. 
	In the notation of Crespo \textit{et al.} (2013)~\cite{crespo2013competing}, these are $J_1$: interactions between edge-sharing MnO$_6$ octahedra where the Mn-Mn distance is parallel with the $c$-axis and the direction of the tunnels; $J_2$: a second edge-sharing interaction between MnO$_6$ octahedra where the Mn-Mn interaction is approximately 29$^\circ$ from the $ab$-plane; and $J_3$: corner-sharing MnO$_6$ octahedra, which are separated from one another by bond-angle $\theta_\mathrm{Mn-O-Mn}$$\approx$130$^\circ$. 
	These are labeled schematically in Figure~\ref{fig_struc}(b).

	The magnetic properties of cryptomelane are found to be dependent on the K-stoichiometry, $x$, with quite different temperature-dependent magnetic behavior in the range $0.72<x<1$ compared with $1<x<2$. 
	For the lower $x$-range, short-range ferromagnetic or ferrimagnetic correlations occur along with formation of a spin glass, with a transition reported in the range $33.1 < T_\mathrm{g} \mathrm{(K)} < 50$~\cite{suib1994magnetic,luo2009spin,luo2010tuning}. 
	We are primarily interested in the higher-$x$ compositional range, where there are 3 reported transitions observed in magnetometry, and a lack of consensus on their origin~\cite{sato1997magnetism,sato1999charge,luo2010tuning,tseng2015magnetic,barudvzija2016structural,barudvzija2020magnetic}. 
	These three transitions are denoted, in descending order of temperature, as $T_1$, $T_2$, and $T_3$, in this and other works~\cite{sato1999charge}.
	The majority of studies of the magnetic structure of cryptomelane with $x \ge 1$ report two magnetic transitions, one at $T_2 \approx 55$\,K and another in the range $T_3\approx20-25$\,K. 
	In some works, a higher temperature transition observed by magnetization and electrical transport measurements, occurring in the range $180<T_1\mathrm{(K)}<250$, is assigned to charge ordering of Mn$^{3+}$ and Mn$^{4+}$~\cite{sato1999charge,luo2010tuning}. 
    However, we note that similar transitions are observed in titanate (IV) hollandites~\cite{Cheary1992Electron16,carter2005universally,larson2017metal} which do not have mixed-metal valence, suggesting a more likely origin is in cation order within the nanotunnels. 
	No associated superstructure has thus far been reported by diffraction for cryptomelane, unlike in other hollandites~\cite{Cheary1992Electron16,carter2005universally}. 
    
	Based on magnetic susceptibility measurements, the transition at $T_2 \approx 55$\,K is described as having non-collinear antiferromagnetic ordering, helical magnetism or weak ferromagnetism~\cite{sato1999charge,luo2010tuning,tseng2015magnetic,barudvzija2016structural}. 
	Studies on single crystals~\cite{sato1997magnetism,sato1999charge} involving isothermal magnetization measurements show that there is significant anisotropy and irreversibility when the field is applied perpendicular to the tunnels, indicating a net moment in the $ab$-plane. 
	
	Below the lower temperature transition $T_3 \approx 20$\,K, some studies report formation of a spin glass-like state~\cite{barudvzija2016structural,barudvzija2020magnetic}. 
	This is based on observed frequency dependence of transition temperature in AC susceptibility and divergence between temperature-dependent susceptibility during cooling in the presence/absence of a magnetic field. 
	Additionally, the spin glass-like state is believed to relate to observed magnetic memory behavior~\cite{barudvzija2016structural,barudvzija2020magnetic}, in which holding the material at a particular temperature in a magnetic field during cooling leads to features at that temperature during subsequent heating~\cite{barudvzija2020magnetic}. 
	A prior neutron diffraction study on a sample of K$_{1.72}$Mn$_8$O$_{16}$ observed a different set of magnetic Bragg peaks in each temperature regime~\cite{larson2017frustrated_chapter}; the authors did not solve the magnetic structure, but do suggest an incommensurate magnetic structure in the higher temperature regime and a commensurate magnetic structure below $\sim$25\,K. 
	Besides this, no neutron diffraction experiments have so far explored the long-range magnetic order in cryptomelane of any stoichiometry, although a study on the Ba-analog~\cite{larson2015inducing} has found modulated magnetism, commensurate with the unit cell.

	In this work, we show that there are both magnetic and structural incommensurate modulations observable by diffraction in cryptomelane with average Mn oxidation state of approximately +3.75. 
    The structural modulation is observed below $T_1 \approx 185$\,K, concomitant with a monoclinic distortion from the tetragonal aristotype, and the magnetic modulation is observed below $T_2 \approx 54.5$\,K.
	This is the first observation of incommensurate structural modulations in cryptomelane, and we find that the structural modulation at $T<T_1$ relates to stoichiometry via the same mathematical relation as other hollandites~\cite{carter2005universally}. 
    While it is not the first observation of incommensurate magnetic modulations~\cite{larson2017frustrated_chapter}, it is the first to propose a propagation vector and irreducible representation and show that it fits the data. 
	Our results are compared to theoretical simulations by Mandal \textit{et al}~\cite{mandal2014incommensurate}, with the periodicity of the incommensurate magnetic modulation within the range of their predictions; this periodicity is also approximately a factor of 2 that of the structural modulation, reminiscent of the ``helical ferrimagnetism" proposed by Sato \textit{et al}~\cite{sato1997magnetism,sato1999charge}, though not entirely consistent with the mechanism as they formulated it. 
	We also observe the existence of a set of magnetic Bragg peaks, below the transition at $T_3 \approx 24$\,K, albeit at different positions compared with above $T_3$, despite previous suggestions that $T_3$ is a glass transition. 
    However, these peaks cannot be indexed to any simple propagation vector, or even a simple commensurate/incommensurate pair, and we believe a single-crystal neutron study would be needed to determine the magnetic behavior below $T_3$. 

	\begin{figure}[t]
		\includegraphics[width=0.5\linewidth]{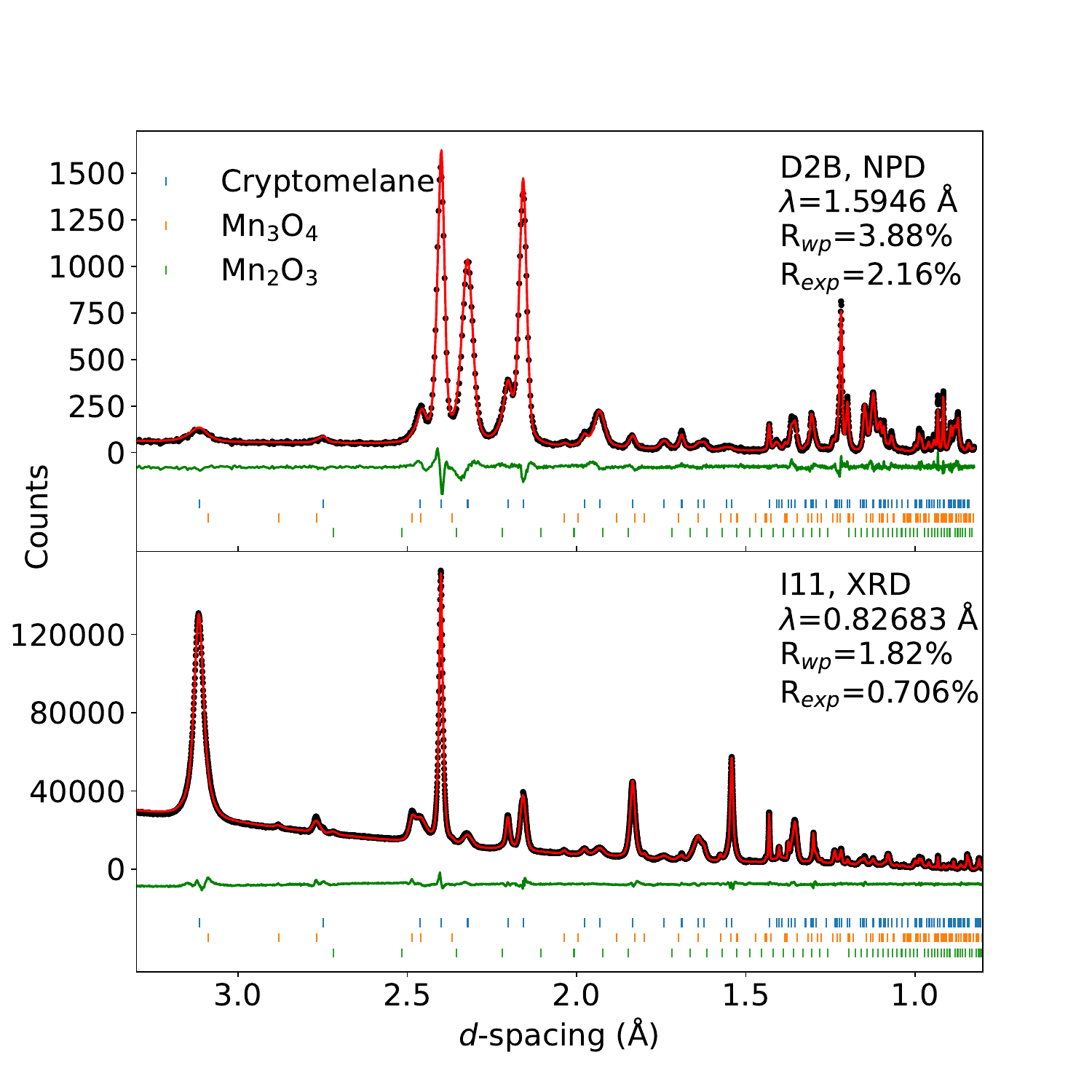}
		\caption{
			\label{fig_refinements}
			Experimental (black points) and calculated (red line, modeled by combined Rietveld refinement) diffraction data for KMO-S1 at room temperature for synchrotron XRD (bottom) and neutron diffraction (top). 
			The difference between the model and data is shown in green. 
			Blue tick-marks indicate nuclear Bragg peaks due to cryptomelane K$_{1.448(3)}$Mn$_8$O$_{16}$, and orange and green tick-marks indicate nuclear Bragg peaks due to the Mn$_3$O$_4$ (2.8(3)wt\%) and Mn$_2$O$_3$ (0.049(5)wt\%) impurities respectively. 
			The R$_{wp}$ and R$_{exp}$ of the fit to each dataset are shown in the figure panels, and the values for the total refinement were 1.87\% and 0.750\% respectively (to 3 significant figures).
		}
	\end{figure}

	\section{\label{sec:level2}Methods}

	\paragraph{Sample preparation.}

	Cryptomelane was prepared, in a quantity of several grams, using a solid-state synthesis route with a target composition of K$_{1.33}$Mn$_8$O$_{16}$. 
	The precursors of KNO$_3$ and MnCO$_3$ (99.994\% and 99.9\% purity respectively; both from Alfa Aesar) were mixed thoroughly using an agate mortar and pestle, with a 5\% molar excess of KNO$_3$ to account for K-loss during high-temperature calcination, corresponding to an actual precursor molar ratio of 1.3965:8 KNO$_3$:MnCO$_3$. 
	Planetary mixing in ethanol was carried out for $\sim$15 minutes. 
	The resultant powder was annealed in air at 500\,$^\circ$C for $\sim$4 hours, with a heating rate of 5\,$^\circ$C min$^{-1}$. 
	Samples were cooled to room temperature by switching off the furnace.

		KMO-S1 and KMO-S2 denote two batches of cryptomelane prepared independently using this route; KMO-S1 was used for all measurements described in the main text except the variable-temperature synchrotron diffraction, for which KMO-S2 was used (see also Table~S1). 
		Combined neutron and synchrotron refinement allowed K occupancy, and hence composition, to be determined precisely for KMO-S1 (K$_{1.448(3)}$Mn$_8$O$_{16}$), near-identical to the K content based on impurity fractions (K$_{1.447(5)}$Mn$_8$O$_{16}$). 
        An equivalent refinement was not possible for KMO-S2; however, despite closely comparable lattice parameters of the two batches (within 2$\times10^{-3}$\,\AA{} of one another, as shown in Table~S2) and essentially identical magnetic properties, Mn impurity content is slightly larger for KMO-S2, implying a higher K occupancy of around K$_{1.488(2)}$Mn$_8$O$_{16}$ (Table~S2).

	\paragraph{Magnetization measurements.}

	A Quantum Design Magnetic Properties Measurement System (MPMS-3) was used for DC susceptibility and isothermal magnetization measurements. 
	Measurements were taken during warming after cooling in zero field (ZFC-W) and in the measuring field (FC-W), where the measuring field was 200\,Oe. 
	Isothermal magnetization measurements were performed in an external magnetic field between $\pm70$\,kOe.

	A Quantum Design 9\,T Physical Properties Measurement System (PPMS) using the ACMS-II option was used for AC susceptibility. 
	Measurements were taken with an AC driving field of 10\,Oe in a static field of 10\,Oe at 501\,Hz, 1001\,Hz, 1501\,Hz, 2002\,Hz, and 4004\,Hz.

	\paragraph{Heat capacity.}

	A Quantum Design PPMS was used for heat capacity measurements. 
	As is common for thermally insulating samples~\cite{melot2009large,mukherjee2017enhanced,kelly2020structure,kelly2022magnetism,koskelo2022free} the sample was mixed with silver powder (Alfa Aesar, 99.99\%, -635 mesh) in a 1:1 mass ratio to increase the conductivity; this was accounted for using tabulated temperature-dependent heat capacity of silver from Ref.~\citen{smith1995low}. 
	Apiezon N grease was used to provide thermal contact between the sample platform and the pellet. 
	The lattice contribution to the heat capacity was estimated and accounted for using a Debye temperature~\cite{debye1912theorie} $T_\mathrm{D} = 301.5(11)$\,K, obtained from a two-parameter fit (the Debye temperature $T_\mathrm{D}$ and an overall multiplicative scale factor only) using \textsc{SciPy}~\cite{2020SciPy-NMeth} (see Figure~S2 of Supplementary Information).
	Uncertainty in $T_\mathrm{D}$ was obtained from the diagonal elements of the fit's parameter covariance matrix, and propagated to the background-subtracted heat capacity which accounts for the correlation between $T_\mathrm{D}$ and the scale factor.

	\paragraph{Diffraction.}

	Synchrotron X-ray Diffraction (XRD) was performed using the I11 instrument~\cite{thompson2009beamline,thompson2011fast} at Diamond Light Source. 
	As the measurements were taken during different runs, $\lambda= 0.82683$\, \AA{} for room temperature; $\lambda= 0.825223 $\,\AA{} for low temperatures from 300\,K to 100\,K. 
	Synchrotron diffraction at I11 used the Mythen-2 position-sensitive detector, with a collection time of $\sim$20\,seconds per diffraction pattern. 
	The sample was contained in a 0.5\,mm diameter glass capillary, sealed with epoxy (Loctite Double Bubble). 
	For the low-temperature synchrotron X-ray diffraction between 100\,K and 300\,K, a discrete temperature spacing of 1\,K was used.
	Low-temperature laboratory XRD below 100\,K was performed using a Bruker D8 Discover diffractometer (Cu\,K$\alpha$; $\lambda =1.541$\,\AA), with an Oxford Cryosystems PheniX closed-cycle refrigerator stage, using a two-stage Gifford–McMahon cryocooler.

	Constant-wavelength neutron diffraction experiments were performed at the Institut Laue-Langevin. 
	Measurements were performed on the high-resolution powder diffractometer D2B (room-temperature; $\lambda= 1.5946 $\,\AA) and the high-intensity powder diffractometer D20 (2.5\,K, 40\,K, 44\,K, 47\,K, and 60\,K; $\lambda= 2.4108 $\,\AA). 
	The neutron diffraction was performed on an aliquot of the same sample batch, KMO-S1, as the room-temperature synchrotron diffraction data used for combined Rietveld analysis. 
	The variable-temperature synchrotron diffraction data discussed later were collected on a separately-synthesized batch, KMO-S2, and combined neutron and X-ray diffraction were not performed for KMO-S2. 
	

	\paragraph{Diffraction analysis.}

	Structural diffraction data were analyzed using the software package \textsc{Topas-Academic}~\cite{coelho2018topas} using Rietveld refinement~\cite{rietveld1969profile}. 
	The room-temperature data were analyzed by combined refinement of the room-temperature I11 and D2B data, whereas the variable-temperature synchrotron diffraction data were analyzed by sequential Rietveld refinement in which each diffraction pattern is analyzed in ascending temperature order, with the starting parameters of a refinement being the refined parameters from the previous diffraction pattern. 
	The background was fitted by a Chebyshev polynomial (order 15 and 6 for synchrotron data and neutron data respectively in the room-temperature refinement; order 20 in the sequential refinement of synchrotron XRD). 
	A fixed Thompson-Cox-Hastings pseudo-Voigt peak-shape refined from standard Si data was used~\cite{thompson1987rietveld} for the instrumental contribution to the synchrotron data, and the sample contribution to the peak-shape was accounted for using Stephens' function for anisotropic microstrain broadening~\cite{stephens1999phenomenological} convoluted with a Lorentzian particle size broadening function. 
	The neutron peak shape was modeled using the Lorentzian and Gaussian particle size functions, and also Stephens' function~\cite{stephens1999phenomenological}. 
	Peak asymmetry was accounted for using a simple axial divergence model. 
	All atomic positions, lattice parameters, and atomic displacement parameters were refined within symmetry constraints. 
	Anisotropic atomic displacement parameters (ADPs)~\cite{peterse1966anisotropic} were refined for K and O, whereas isotropic thermal displacement parameters were used for Mn. 
	Mn occupancies were fixed to 100\% while all K and O occupancies were initially refined. 
	The O2 site was fully occupied within error, and so was fixed at 100\% for the final refinement.

	For combined refinement of synchrotron X-ray and neutron diffraction data on KMO-S1, a relative weighting scheme was established such that both datasets contributed equally to the refinement despite the inequivalent signal-to-noise. 
	This involved varying relative weightings by trial and error until the total $\chi^2$ was approximately equal to the average of the individual $\chi^2$ values for each dataset.

	Magnetic diffraction analysis was performed using \textsc{FullProf}~\cite{rodriguez2001fullprof}. 
	Diffraction difference patterns were calculated by subtracting the neutron 60\,K dataset from lower-temperature diffraction patterns using LAMP~\cite{richard1996analysis}; we note that due to the very small lattice parameter shifts it was not necessary to correct for thermal expansion. 
	Candidate magnetic propagation vectors ($\vec{k}_\mathrm{mag}$) were found using K Search, a tool in \textsc{FullProf}, using the position of peaks in the difference pattern. 
	Precise values of parameters in $\vec{k}_\mathrm{mag}$ were refined against the difference pattern by Le Bail refinement~\cite{le1988ab}. 
	\textsc{BASIREPS}~\cite{rodriguez2010program,ritter2011neutrons} was used to calculate the Irreducible Representations (IRs) for a given $\vec{k}_\mathrm{mag}$ (see Section 7 of Supplementary Information) and their basis vectors (BVs). 

	In \textsc{FullProf}, the March-Dollase model~\cite{dollase1986correction} was used to account for preferred orientation in the sample used for low-temperature neutron diffraction due to slightly non-spherical particle shape (see Figure~S1 of Supplementary Information).

	\section{\label{sec:level4}Results}

	\subsection{Structural characterization}
	\label{Structural_section}

	\begin{table}[htbp]\small
		\caption{
			\label{table_refinements_tetragonal_RT}
			The crystallographic parameters of cryptomelane KMO-S1 at room temperature, as obtained by a combined Rietveld refinement of XRD data from the I11 instrument and neutron diffraction using the D2B instrument. 
            Note that for K, $U_{11}=U_{22}$ by symmetry. 
			Mn and O2 occupancies were fixed at 1. 
		}
		\begin{tabular}{|ll|}
			\hline
			\multicolumn{1}{|l|}{Space group} & \multicolumn{1}{c|}{\textit{I4/m}}  \\ \hline
			\multicolumn{1}{|l|}{$a=b$ / \AA} & \multicolumn{1}{l|}{9.84942(9)} \\ \hline
			\multicolumn{1}{|l|}{$c$ / \AA} & \multicolumn{1}{l|}{2.862033(17)} \\ \hline
			\multicolumn{1}{|l|}{$V$ / \AA$^3$} & \multicolumn{1}{l|}{277.649(6)} \\ \hline
			\multicolumn{1}{|l|}{$\chi^2$} & \multicolumn{1}{l|}{6.21}  \\ \hline
			\multicolumn{1}{|l|}{R$_{wp}$} & \multicolumn{1}{l|}{1.87\%} \\ \hline
			\multicolumn{1}{|l|}{R$_{exp}$} & \multicolumn{1}{l|}{0.750\%} \\ \hline
			\multicolumn{2}{|c|}{K at 4\textit{e} (0, 0, \textit{z})} \\ \hline   
			\multicolumn{1}{|l|}{$z$(K)} & \multicolumn{1}{l|}{0.3559(11)} \\ \hline
			\multicolumn{1}{|l|}{$U_{11}$ / \AA$^2$} & \multicolumn{1}{l|}{0.0259(11)} \\ \hline
			\multicolumn{1}{|l|}{$U_{33}$ / \AA$^2$} & \multicolumn{1}{l|}{0.093(4)} \\ \hline
			\multicolumn{1}{|l|}{occupancy(K)} & \multicolumn{1}{l|}{0.3621(8)} \\ \hline
			\multicolumn{2}{|c|}{Mn at 8\textit{h} (\textit{x}, \textit{y}, 0)} \\ \hline
			\multicolumn{1}{|l|}{$x$(Mn)} & \multicolumn{1}{l|}{0.34939(4)}  \\ \hline
			\multicolumn{1}{|l|}{$y$(Mn)} & \multicolumn{1}{l|}{0.16575(5)}  \\ \hline
			\multicolumn{1}{|l|}{$U_{11}$(Mn) / \AA$^2$} & \multicolumn{1}{l|}{0.0086(4)} \\ \hline
			\multicolumn{1}{|l|}{$U_{22}$(Mn) / \AA$^2$} & \multicolumn{1}{l|}{0.0069(4)} \\ \hline
			\multicolumn{1}{|l|}{$U_{33}$(Mn) / \AA$^2$} & \multicolumn{1}{l|}{0.0061(3)} \\ \hline
			\multicolumn{1}{|l|}{$U_{12}$(Mn) / \AA$^2$} & \multicolumn{1}{l|}{0.0027(3)} \\ \hline
			\multicolumn{2}{|c|}{O at 8\textit{h} (\textit{x}, \textit{y}, 0)} \\ \hline
			\multicolumn{1}{|l|}{$x$(O1)} & \multicolumn{1}{l|}{0.15315(12)}  \\ \hline
			\multicolumn{1}{|l|}{$y$(O1)} & \multicolumn{1}{l|}{0.20283(10)} \\ \hline
			\multicolumn{1}{|l|}{$U_{11}$(O1) / \AA$^2$} & \multicolumn{1}{l|}{0.0103(9)} \\ \hline
			\multicolumn{1}{|l|}{$U_{22}$(O1) / \AA$^2$} & \multicolumn{1}{l|}{0.0158(6)} \\ \hline
			\multicolumn{1}{|l|}{$U_{33}$(O1) / \AA$^2$} & \multicolumn{1}{l|}{0.0028(6)} \\ \hline
			\multicolumn{1}{|l|}{$U_{12}$(O1) / \AA$^2$} & \multicolumn{1}{l|}{0.0000(7)} \\ \hline	
			\multicolumn{1}{|l|}{occupancy(O1)} & \multicolumn{1}{l|}{0.9801(12)} \\ \hline
			\multicolumn{1}{|l|}{$x$(O2)} & \multicolumn{1}{l|}{0.54063(12)}  \\ \hline
			\multicolumn{1}{|l|}{$y$(O2)} & \multicolumn{1}{l|}{0.16551(13)} \\ \hline
			\multicolumn{1}{|l|}{$U_{11}$(O2) / \AA$^2$} & \multicolumn{1}{l|}{0.0055(8)} \\ \hline
			\multicolumn{1}{|l|}{$U_{22}$(O2) / \AA$^2$} & \multicolumn{1}{l|}{0.0121(8)} \\ \hline
			\multicolumn{1}{|l|}{$U_{33}$(O2) / \AA$^2$} & \multicolumn{1}{l|}{0.0030(6)} \\ \hline
			\multicolumn{1}{|l|}{$U_{12}$(O2) / \AA$^2$} & \multicolumn{1}{l|}{0.0005(7)} \\ \hline
			\multicolumn{2}{|c|}{Bond lengths} \\ \hline 
			\multicolumn{1}{|l|}{Mn-Mn ($J_1$; x 2) / \AA} & \multicolumn{1}{l|}{2.862033(17)} \\ \hline
			\multicolumn{1}{|l|}{Mn-Mn ($J_2$; x 2) / \AA} & \multicolumn{1}{l|}{2.9385(8)} \\ \hline
			\multicolumn{1}{|l|}{Mn-Mn ($J_3$; x 4) / \AA} & \multicolumn{1}{l|}{3.4321(6)} \\ \hline
			\multicolumn{1}{|l|}{Mn-O1 (x 1) / \AA} & \multicolumn{1}{l|}{1.967(1)} \\ \hline
			\multicolumn{1}{|l|}{Mn-O1 (x 2) / \AA} & \multicolumn{1}{l|}{1.9297(7)} \\ \hline
			\multicolumn{1}{|l|}{Mn-O2 (x 1) / \AA} & \multicolumn{1}{l|}{1.884(1)} \\ \hline
			\multicolumn{1}{|l|}{Mn-O2 (x 2) / \AA} & \multicolumn{1}{l|}{1.8942(8)} \\ \hline
			\multicolumn{2}{|c|}{Mn-O-Mn bond angles} \\ \hline
			\multicolumn{1}{|l|}{Mn-O1-Mn / $^\circ$} & \multicolumn{1}{l|}{97.89(5)} \\\hline
			\multicolumn{1}{|l|}{Mn-O2-Mn / $^\circ$} & \multicolumn{1}{l|}{98.13(6)} \\
			\hline
		\end{tabular}
	\end{table}

	The room temperature structure of cryptomelane, K$_x$Mn$_8$O$_{16}$, was investigated by neutron and X-ray diffraction on the D2B instrument at the ILL and the I11 instrument at Diamond Light Source, respectively. 
	Combined Rietveld refinement was carried out with the previously-reported $I4/m$ (space group \#87) structure~\cite{vicat1986structure}. 
	In this tetragonal symmetry structure, there is a single Mn site, 8\textit{h} (\textit{x}, \textit{y}, 0), and two inequivalent oxygen sites both at 8\textit{h} (\textit{x}, \textit{y}, 0). 
	K is located within the tunnels in a partially occupied site, 4\textit{e} (0, 0, \textit{z}).

	Structural parameters and refinement R-factors from a combined Rietveld refinement on the room-temperature X-ray on KMO-S1 and neutron diffraction data are shown in Table~\ref{table_refinements_tetragonal_RT}, with the refined and calculated profile shown in Figure~\ref{fig_refinements}. 
	These measurements indicate the presence of a 2.8(3)\% (by weight) Mn$_3$O$_4$ impurity phase ($I4_1/amd$ space group, \#141). 
	There are also a number of very small peaks, visible only in the synchrotron diffraction data, which are consistent with the presence of trace amounts of Mn$_2$O$_3$ ($Ia\bar{3}$; \#206); this impurity phase refines to 0.049(5)\% (by weight). 
	Such Mn$_x$O$_{x+1}$ impurities are a common occurrence in magnetic studies of manganese-containing materials~\cite{bocarsly2019deciphering,przenioslo1999magnetic,munoz2008ferromagnetic}. 
	Refinement of K site occupancy gives $x=1.448(3)$ in K$_x$Mn$_8$O$_{16}$ for KMO-S1. 
    The excess K, as compared with the target composition of K$_{1.33}$Mn$_8$O$_{16}$, is a consequence of the formation of the K-free Mn$_3$O$_4$/Mn$_2$O$_3$ impurity phases and the use of excess K during synthesis; we can be confident of this based on a near-identical $x=1.447(5)$ calculated by determining remaining K based on refined weight percents of impurity phases [Table~S2]. 
    Using this same approach based on weight percents of impurities, we estimate KMO-S2 to have a slightly higher K content, with $x=1.488(2)$. 
	The room-temperature Rietveld-refined crystallographic model of both KMO-S1 and KMO-S2 is consistent with the reported crystal structure for cryptomelane. 

	In the combined Rietveld fit on KMO-S1 , anisotropic strain was implemented using Stephens' function~\cite{stephens1999phenomenological}. 
	Refined strain parameters obtained from the synchrotron data, $S_{004}=2970(60)$ and $S_{400}=1637(9)$, show the anisotropic nature of this strain. 
	Anisotropic strain likely comes from the tunnel dimensions in hollandites being highly dependent on the occupancy of the large cation in the tunnels, as shown by previous works on cryptomelane~\cite{poyraz2017synthesis} and its silver analog Ag$_x$Mn$_8$O$_{16}$ (1.14$\leq x\leq$1.66)~\cite{brady2018effect}. 
	The strain is likely in part due to local variations in the K content and resultant variations on Mn oxidation state (due to the requirement for local charge balancing) leading to local differences. 

	Anisotropic ADPs for the K$^+$ cations in KMO-S1 refined to be highly anisotropic, with the $U_{33} = 0.093(4)$\,\AA{}$^2$ several times larger than $U_{22}=U_{11}=0.0259(11)$\,\AA{}$^2$, as shown in Figure~\ref{fig_struc}(a). 
	This is consistent with previous work~\cite{boschetti2023cryptomelane} which showed that the anisotropic K$^{+}$ ADPs could also be modeled using a split-site approach. 
	The high degree of anisotropy in the K$^{+}$ ADPs could be due to static disorder from K$^{+}$ ions occupying a range of $z$ positions in the 4\textit{e} (0, 0, \textit{z}) sites, for instance due to different K$^+$ positioning in proximity to a K$^+$ cation or vacancy.
	Alternatively, there may be dynamic disorder from highly mobile K$^{+}$ hopping between adjacent sites. 

	The ADPs for the two oxygen sites show different behavior, likely resulting from the different degrees of freedom due to this MnO$_6$ connectivity. 
	The local environments of the two oxygen sites, O1 and O2, are different. O1 oxygen sites share bonds with K and Mn, and occur as the shared vertices of edge-sharing MnO$_6$ octahedra, whereas O2 oxygen sites are further from K than O1, and are the shared vertices of three corner-sharing MnO$_6$ octahedra. 
	Refinement of site occupancies for oxygen indicated oxygen vacancies on one of the two O sites, with a refined occupancy on the O1 site of 0.9801(12). 
	This corresponds to K$_{1.448(3)}$Mn$_8$O$_{15.841(9)}$, and from this stoichiometry we calculate an average Mn oxidation of +3.779(2). 
	For simplicity, we will use the nominal oxygen content in all further discussion.

	Another way to determine oxidation state is using the empirical bond valence sum (BVS) relation between bond length and oxidation state~\cite{brown1985bond}, which was previously applied to Ba$_{1.2}$Mn$_8$O$_{16}$~\cite{larson2015inducing}. 
	For the MnO$_6$ octahedra in this study, we can obtain Mn valence $V$ using the following equation:

	\begin{equation}\label{BVS_equation}
		V = \sum_{i=1}^6 \exp{\left[\frac{r_0-r_i}{B}\right]}
	\end{equation}

	where the empirical parameters $r_0=1.762$\,\AA{} and $B=0.34$\,\AA{} for Mn$^{4+}$ (taken from Ref.~\citen{urusov2006problem}), and $r_i$ are the Mn-O bond lengths. 
	Using this approach, we obtain an average Mn oxidation state of +3.826(4). 
	This is broadly consistent with the value obtained based on composition (+3.79(2)) given the precision that can be expected for this type of calculation.

	The MnO$_6$ octahedra are all identical in the average $I4/m$ structure. 
	They exhibit four different Mn-O bond lengths between 1.8837(13)\,\AA{} and 1.967(1)\,\AA. 
	It is possible that some portion of the bond length distortion may be in part the result of averaging multivalent MnO$_6$ octahedra, some of which contain Jahn--Teller-active Mn$^{3+}$, but the competing geometric interactions on each octahedron add complexity. 
	These bond length values and multiplicity does not lend itself to simple deconvolution of the $Q_2$ and $Q_3$ van Vleck modes with $E_g$ symmetry typical of a Jahn--Teller distortion~\cite{van1939jahn,naglecocco2024van}, unlike for instance the perovskite LaMnO$_3$~\cite{rodriguez1998neutron} which has three unique bond lengths each mutually perpendicular, and so possible signatures of a Jahn--Teller distortion cannot be determined. 
	The bond-length distortion index~\cite{baur1974geometry}, $D$, is a physical parameter which is often used~\cite{lawler2021decoupling,kimber2012charge,nagle2022pressure,genreith2024jahn} to quantify the degree of bond length distortion of coordinated species around a central ion. 
	It is defined as:

	\begin{equation}
	\label{BLDI_equation}
		D=\frac{1}{n} \sum_{i=1}^n \frac{|l_i - l_\mathrm{av}|}{l_\mathrm{av}}
	\end{equation}

	where $l_i$ is the distance between the core ion and the $i$th coordinated ion, $l_\mathrm{av}$ is the average of all the distances between the core ion and coordinated ions, and $n=6$ for an octahedron. 
	We here calculate a $D=0.01320(12)$, which is similar to that previously reported for mixed Mn$^{3+/4+}$O$_6$ octahedra (with an average oxidation state of $\sim$3.42+) in Pb$_3$Mn$_7$O$_{15}$~\cite{kimber2012charge}. 
	We can compare the absence of a cooperative Jahn--Teller distortion to mixed-valent Mn$^{3+/4+}$ system LiMn$_2$O$_4$ spinel which, even with a much larger Mn$^{3+}$ fraction ($\sim$50\%, corresponding to a mean valence +3.5) than the cryptomelane sample studied here, does not exhibit Jahn--Teller order.
	However, Jahn--Teller order onsets on electrochemical lithiation towards Li$_2$Mn$_2$O$_4$ with a transition from cubic to tetragonal symmetry~\cite{falqueto2023unveiling} as mean Mn valence approaches +3. 
	This suggests that, with a mean Mn valence of +3.779(2), the Mn$^{3+}$ fraction in our sample of K$_{1.448(3)}$Mn$_8$O$_{16}$ is simply too low to drive long-range cooperative Jahn--Teller order. 

	\subsection{Low-temperature synchrotron diffraction}
	\label{lowT_diffraction_section}

	\begin{figure}[t]
		\includegraphics[width=\linewidth]{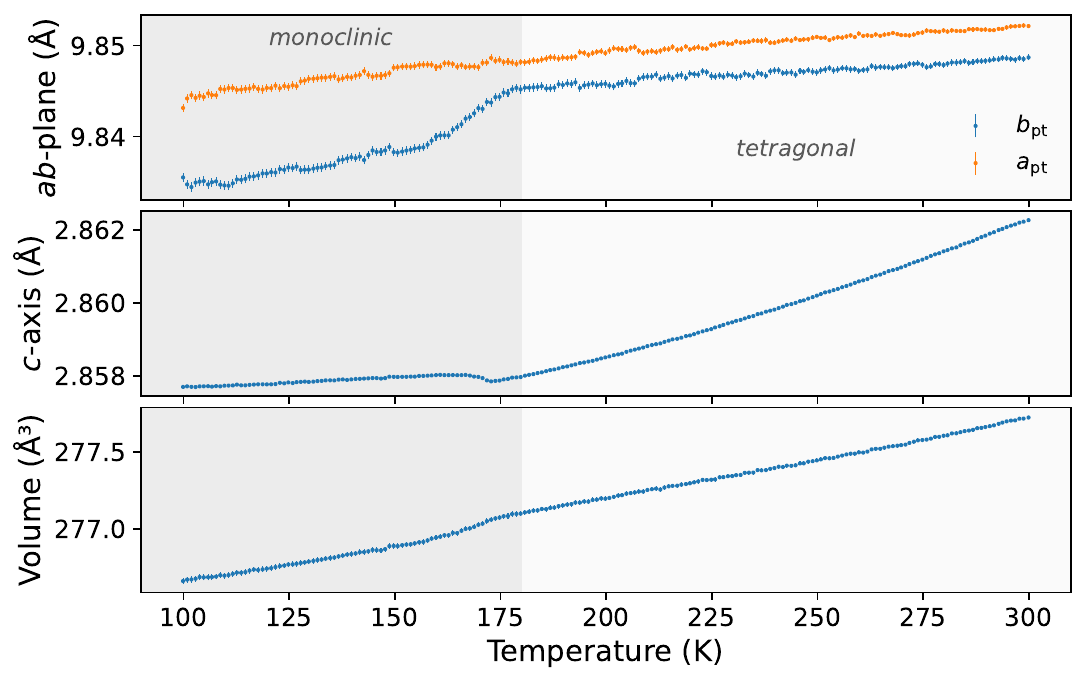}
		\caption{
			\label{Pseudo-tetragonal_Distances_vs_Temperature.pdf}
			Top: Pseudo-tetragonal lattice parameters, $a_\mathrm{pt}$ and $b_\mathrm{pt}$ (Eqs~\ref{eq_pt_a} and \ref{eq_pt_b}); middle: lattice parameter $c$ parallel to nanotunnels; bottom: unit cell volume, all as a function of temperature. 
			Data obtained from sequential Rietveld refinement of synchrotron diffraction data of KMO-S2 using the $A2/m$ space group; we note that above the $T_1$ transition the tetragonal cell is the better descriptor of the average structure [Figure~S24]. 
			Temperature dependence of original monoclinic lattice parameters is shown in Figure~S12. 
		}
	\end{figure}

	Variable-temperature synchrotron diffraction was performed on KMO-S2 during cooling from 300\,K to 100\,K to examine the low-temperature crystal structure of cryptomelane. 
	A tetragonal symmetry was observed on cooling from room-temperature to $T_1$, although a continuous increase in anisotropic peak broadening can be seen [Figure~S16]. 
	$T_1$ was observed to occur at $\sim$175\,K, below which two significant changes associated with the cryptomelane phase occurred. 
	Firstly, diffraction peaks began to broaden and even split [Figures~S10(b) and S9(e)], suggesting that the average structure is lowered due to a subtle monoclinic distortion, similar to that reported by Boschetti \textit{et al.}~\cite{boschetti2023cryptomelane} for cryptomelane; such a first-order transition has been reported previously for other hollandites~\cite{cheary1986analysis,cheary1990structural}. 
	Secondly, a series of weak superlattice peaks appeared, consistent with a non-magnetic structural modulation [Figure~S9]. 
    We index these to $\vec{k}_\mathrm{struc}=(0,0,k_z)$ where $k_z=0.740834(7)$, via assignment of $(0,\bar{1},1-k_z)$ at $d=7.35$\,\AA{}, $(\bar{1},0,k_z)$ at $d=3.37$\,\AA{}, $(0,2,k_z)$ at $d=3.03$\,\AA{}, and $(\bar{2},0,k_z)$ at $d=2.58$\,\AA{}. 
	As we show in Figure~S27, this structural propagation vector is invariant with temperature. 

	The diffraction data at 100\,K cannot be satisfactorily fit using a tetragonal $I4/m$ cell, and so a monoclinic (space group \#12) cell derived from a distortion of the $I4/m$ structure was used for Rietveld refinement [Figure~S11; Table~S4]. 
    We use the unconventional $A2/m$ setting to keep the tunnel direction along the $c$-axis, consistent with our room-temperature structure. 
	This model does not account for the superlattice peaks, however, which are not fit in the Rietveld refinement. 
	We then performed a sequential Rietveld refinement of the diffraction data to evaluate the nature of the monoclinic distortion and the tetragonal$\leftrightarrow$monoclinic transition. 
	We describe the monoclinic unit cell in terms of pseudo-tetragonal lattice parameters, given by:

	\begin{equation}\label{eq_pt_a}
		a_\mathrm{pt} = a_\mathrm{m}
	\end{equation}

	\begin{equation}\label{eq_pt_b}
		b_\mathrm{pt} = \sqrt{ a_\mathrm{m}^2 + b_\mathrm{m}^2 + 2 a_\mathrm{m} b_\mathrm{m} \cos{\gamma} }
	\end{equation}

	where we use a setting in both cases where $c_\mathrm{m} = c_\mathrm{pt}$ is the direction parallel to the tunnels. 
	Figure~\ref{Pseudo-tetragonal_Distances_vs_Temperature.pdf} shows the temperature dependence of these lattice parameters. 
	Around 100\,K the two pseudo-tetragonal distances are clearly distinct, but they appear to approximately converge at $T_1$; this, along with the monoclinic $\gamma$ converging on the high-symmetry value of 135$^\circ$ [Figure~S12] support the interpretation that $T_1$ is the tetragonal$\leftrightarrow$monoclinic transition. 
	Additionally, we also see a crossover in the magnitude of fit quality (given by $R_\mathrm{wp}$) for sequential Rietveld refinements performed using a tetragonal and monoclinic unit cell [Figure~S24], with the monoclinic model giving a better fit below $T_1$ and the tetragonal model giving a better fit above $T_1$.
	
	Thermal expansion is highly anisotropic in both temperature regimes [Figure~S14]. 
	In the $100$\,K$<T<T_1$ regime, there is very little expansion parallel to the nanotunnels ($\sim0.02$\%) compared with that in the pseudo-tetragonal axes ($\sim0.06$\%). 
	Conversely, in the $T_1<T<300$\,K regime, there is an increase of $\sim0.15$\% in the $c$-axis compared with an increase of $\sim0.03$\% (as a percentage of magnitude at 100\,K) between $T_1$ and 300\,K. 

	In the room-temperature structure there was a single Mn site, whereas in the monoclinic cell there are two Mn sites, both with Wyckoff 4\textit{i}$(x,y,0)$ symmetry: Mn1 and Mn2. 
	We calculate the BVS (using Equation~\ref{BVS_equation}) for each of these sites in Figure~S23, and find radically different valences for each site, calculating 3.37(3) and 4.14(5), respectively for Mn1 and Mn2 at 100\,K. 
	This suggests that Mn2 is entirely in the Mn$^{4+}$ oxidation state, whereas Mn1 consists of a mixture of the Mn$^{4+}$/Mn$^{3+}$ oxidation states. 
	While the distribution of Mn$^{4+}$/Mn$^{3+}$ on the Mn1 site may be entirely random, it is plausible that there may be some incommensurate ordering of Mn$^{4+}$/Mn$^{3+}$ on the Mn1 sublattice giving rise to the weak incommensurate reflections which emerge below $T_1$. 

	As in the room-temperature tetragonal structure, each MnO$_6$ octahedron in the monoclinic structure has four unique Mn-O bond lengths: two are opposite one another, the other two unique bond lengths occur within the same plane, with equal-length bonds occurring 90$^\circ$ from one another. 
	Calculation of bond length distortion index~\cite{baur1974geometry} [Figure~S21] gives a value (0.0283(12)) for Mn1 which is $\sim$2.4 times larger than for Mn2 (0.0116(11)) at 100\,K.
	This is consistent with Mn2 having a non-Jahn--Teller active 4+ oxidation state, while Mn1 appears to be a mixture of Mn$^{3+}$ and Mn$^{4+}$, potentially resulting in Jahn--Teller character. 
	The distortion index value for Mn1 is still small compared with that of Jahn--Teller-distorted NiO$_6$ octahedra in NaNiO$_2$ (0.05463(14))~\cite{nagle2022pressure} or MnO$_6$ octahedra in CuMnO$_2$~\cite{lawler2021decoupling}, but is comparable to mixed-valent MnO$_6$ octahedra in Pb$_3$Mn$_7$O$_{15}$~\cite{kimber2012charge}. 
	We also consider the polar O-O distances in each octahedron, Figure~S22, and see that there are two short and one long distance for each octahedron consistent with a Jahn--Teller distortion localized to the Jahn--Teller-active Mn$^{3+}$ fraction of this mixed-valence site, although for Mn2 a high bond angle variance [Figure~S21] is also a likely reason for the smaller O-O bond length. 
	We therefore conclude that it is plausible that there is a Jahn--Teller distortion on the Mn1 site based on the signatures of a tetragonal octahedral elongation, but the presence of multiple competing geometric constraints on the octahedron and the mixed-valence nature of this site make it impossible to definitively attribute site distortion to the Jahn--Teller effect. 

	\subsection{Magnetometry and heat capacity}
	\label{mag_section}

	\begin{figure*}[t]
		\includegraphics{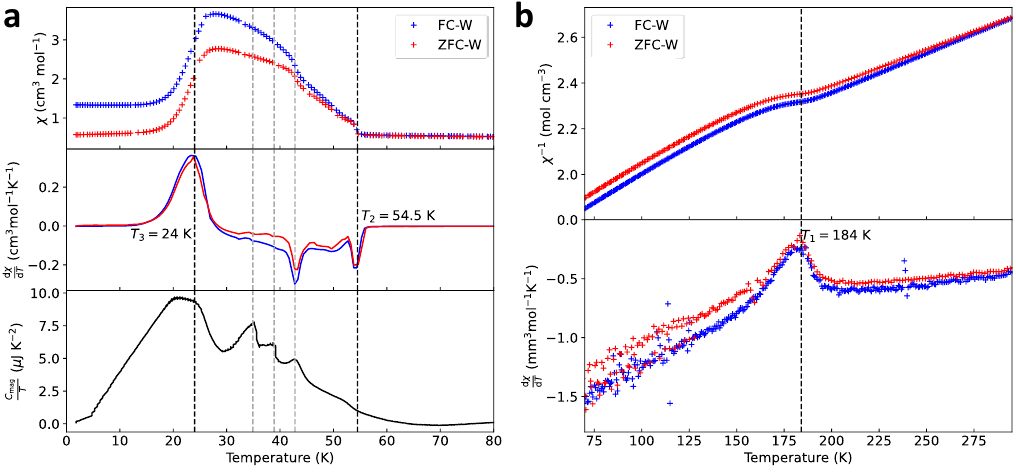}
		\caption{
			\label{fig-mag}
			(a) Magnetic susceptibility $\chi$, $\mathrm{d}\chi/\mathrm{d}T$, and heat capacity as a function of temperature for KMO-S1. 
			(b) Inverse magnetic susceptibility and $\mathrm{d}\chi/\mathrm{d}T$ as a function of temperature in the vicinity of $T_1$. 
			Vertical dashed lines indicate where transition temperature values are assigned; black lines are due to cryptomelane and gray lines are due to Mn$_3$O$_4$~\cite{boucher1971magnetic,boucher1971proprietes,jensen1974magnetic,kemei2014structural}. 
			ZFC-W and FC-W denote zero-field-cooled and field-cooled measurements, measured during warming in a field of 200\,Oe.
		}
	\end{figure*}

	The magnetic transitions in our KMO-S1 cryptomelane sample were studied using DC and AC magnetic susceptibility, heat capacity and isothermal magnetization. 
    We observe each of the three transitions which have previously been observed in K-rich cryptomelane. 
	Here we follow the notation of Ref.~\citen{sato1999charge}, labeling the transitions as $T_1$, $T_2$, and $T_3$ in descending order of temperature.

	Figure~\ref{fig-mag}(a) shows $T_2$ and $T_3$ in the magnetic susceptibility $\chi$, $\frac{\mathrm{d}\chi}{\mathrm{d}T}$, and magnetic heat capacity as a function of temperature, and Figure~\ref{fig-mag}(b) shows $T_1$ in the inverse magnetic susceptibility and $\frac{\mathrm{d}\chi}{\mathrm{d}T}$. 
	The precise values assigned to these transitions are determined from the plot of $\frac{\mathrm{d}\chi}{\mathrm{d}T}$ against temperature: $T_1= 184$\,K, $T_2= 54.5$\,K, and $T_3= 24$\,K. 
	The other transitions observed in the heat capacity and magnetic susceptibility at 34.9\,K, 38.9\,K, and 42.8\,K are all associated with magnetic ordering of Mn$_3$O$_4$ as has been described in the literature~\cite{boucher1971magnetic,boucher1971proprietes,jensen1974magnetic,kemei2014structural}. 
		We also performed Curie-Weiss analysis~\cite{mugiraneza2022tutorial} on our magnetic susceptibility data; given significant limitations with this analysis, we discuss it in detail, along with these limitations, in SI Section~11. 

	\begin{figure}[t]
		\includegraphics{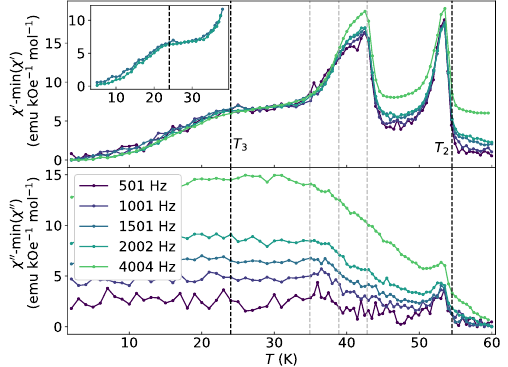}
		\caption{
			\label{AC-fig}
			Variable -temperature and -frequency AC susceptibility measurements on KMO-S1 at 501\,Hz, 1001\,Hz, 1501\,Hz, 2002\,Hz, and 4004\,Hz
			, shown in terms of the real $\chi'$ (top) and imaginary $\chi''$ (bottom) components of the overall AC susceptibility. 
			The inset shows the data for 1501\,Hz and 2002\,Hz only, with a focus on the $T_3$ transition. 
			Positions of vertical lines match those in Figure~\ref{fig-mag}.
		}
	\end{figure}

	\begin{figure}[t]
		\includegraphics[width=87mm]{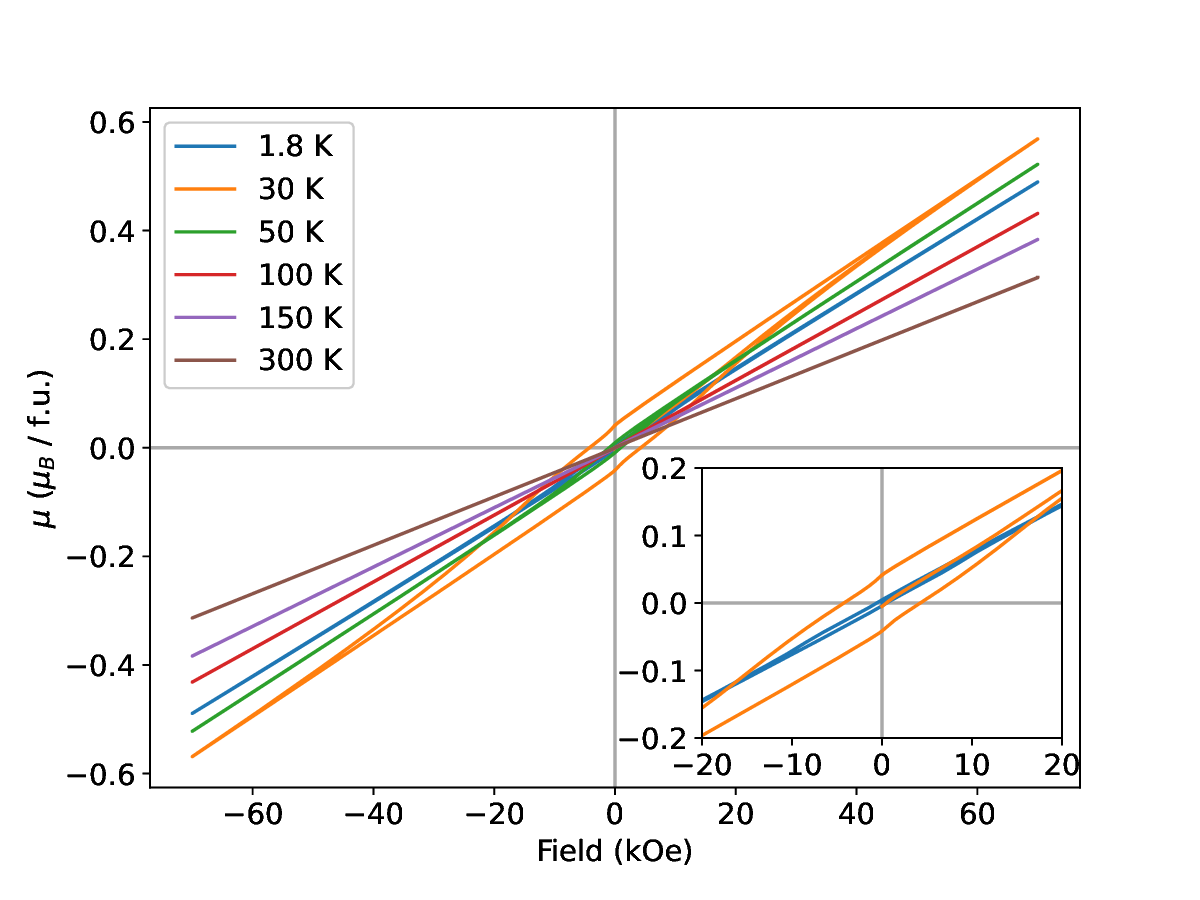}
		\caption{
			\label{MvH-fig}
			Fixed-temperature field-dependent magnetization measurements at 1.8\,K, 30\,K, 50\,K, 100\,K, 150\,K, and 300\,K, showing moment per formula unit of KMO-S1. 
			Inset: 1.8\,K and 30\,K field-dependent magnetization data, zoomed in to highlight the different hysteretic behavior.
		}
	\end{figure}

	The value of $T_1$ at 184\,K, occurs within the range of previously-reported values~\cite{sato1999charge,luo2010tuning}. 
	Reported values of $T_2$ and $T_3$ vary over a broad range. 
	Our values of $T_2=54.5$\,K and $T_3= 24$\,K for KMO-S1 [Table~S10] are slightly higher than most previous studies. 
	Barudžija \textit{et al.} report a $T_2$ transition of either 49\,K or 51\,K (sample-dependent)~\cite{barudvzija2016structural} and Sato \textit{et al.} report 52\,K~\cite{sato1999charge}. 
	For $T_3$, these works report 21\,K or 20\,K, and 20\,K respectively. 
	This may be due to our defining the transition by the maximum in $\frac{\mathrm{d}\chi}{\mathrm{d}T}$, rather than the peak in the $\chi(T)$. 
	Additionally, although there is ostensibly a very similar K$^{+}$ content, our work calculated K$^{+}$ content based on Rietveld refinement whereas the referenced works~\cite{barudvzija2016structural,sato1999charge} calculate K$^{+}$ content based on ratio of precursors. 
	Consequently, there may be a discrepancy in stoichiometry. 
	A theoretical study~\cite{crespo2013competing} shows a dependence of transition temperature on the degree of K$^+$ cation content, and so it is unsurprising that the range of reported temperatures is fairly broad. 
	In our sample, there is divergence between measurements performed after cooling with/without an external magnetic field (FC-C/ZFC-C, respectively), which has previously been interpreted as evidence of spin glass-like ordering~\cite{barudvzija2016structural,barudvzija2020magnetic}; this divergence is present at all measured temperatures (below 300\,K) but becomes significant below $T_2$.

	AC susceptibility measurements between 2\,K and 60\,K are presented in terms of their real $\chi'$ and imaginary $\chi''$ components in Figure~\ref{AC-fig}. 
	$T_2$ manifests as a large feature in both the real and imaginary susceptibility data, peaking around 53.5\,K; this has been reported by Barudžija \textit{et al.}~\cite{barudvzija2016structural,barudvzija2020magnetic} in AC studies on K$_x$MnO$_{2}$ with similar composition. 
	There is a more subtle shoulder for $T_3$ in our AC $\chi$', but no feature can be resolved for $\chi$''. 
	Additionally, the incommensurate ordering of Mn$_3$O$_4$ around 43\,K is present in our data as a large feature in $\chi'$.

	Finally, isothermal DC magnetization measurements are shown in Figure~\ref{MvH-fig}. 
	At 1.8\,K, there is negligible hysteresis, suggesting there is no ferro- or ferri-magnetic behavior for $T<T_3$. 
	However, at 30\,K and, to a lesser extent, 50\,K, there is magnetic hysteresis. 
	This suggests that for $T_3<T<T_2$ there is a ferro- or ferri-magnetic component to the magnetic structure. 
	This is consistent with the previous reports~\cite{sato1999charge,larson2017frustrated_chapter}.

	\subsection{Magnetic ordering from neutron diffraction}

	\begin{figure}[t]
		\includegraphics[width=87mm]{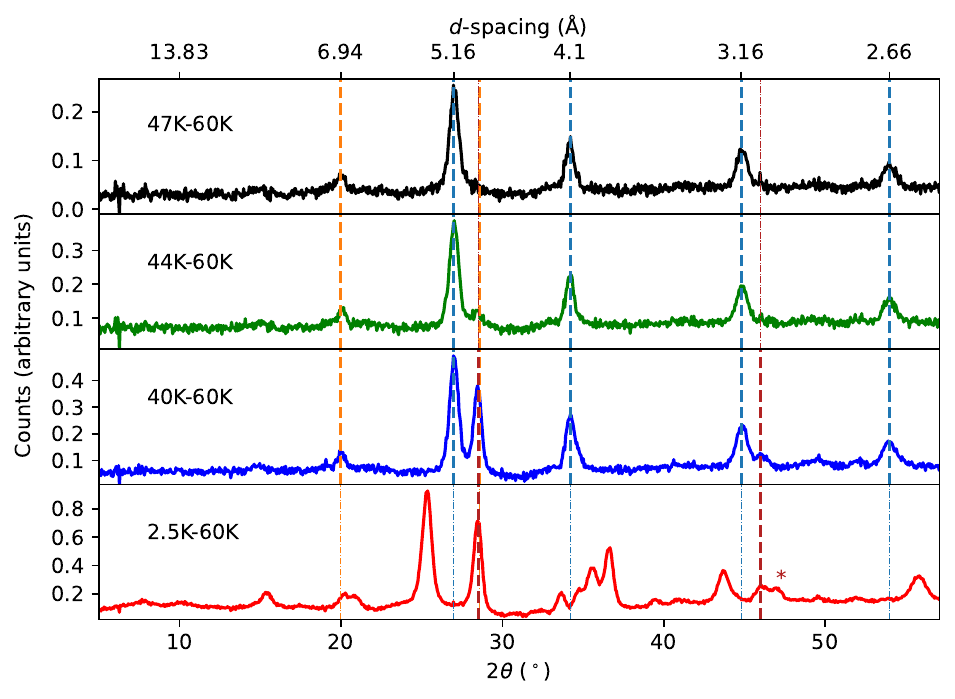}
		\caption{
			\label{diff_patterns}
			Experimental diffraction difference patterns $X$\,K-60\,K ($X$=47\,K, 44\,K, 40\,K, and 2.5\,K) measured on KMO-S1 using the D20 instrument at the ILL. 
			Orange and blue dashed vertical lines indicate peaks associated with the commensurate $\vec{k}_\mathrm{mag}=0$ and incommensurate $\vec{k}_\mathrm{mag}=(0,0,\sim 0.37)$ components for $T_3<T<T_2$, respectively. 
			Dark red dashed vertical lines and asterisks show peaks associated with magnetic ordering in Mn$_3$O$_4$. 
			The vertical lines are thick at the temperatures where the associated peaks are present, and thin at temperatures where they are not.
		}
	\end{figure}

	Variable temperature neutron diffraction was measured on KMO-S1 using the D20 instrument at ILL, Grenoble [SI Figure~S3]. 
	Measurements were collected in the range 2.5\,K to 60\,K in order to study the changes on cooling through $T_2$ and $T_3$. 
	Difference patterns are shown in Figure~\ref{diff_patterns}, in which the diffraction data at 60\,K (with no magnetic diffraction peaks) are subtracted from the lower-temperature diffraction patterns, to allow magnetic Bragg peaks to be readily identified. 
	The changes in pseudo-tetragonal lattice parameters are continuous over the measured temperature points, and all lattice parameters decrease with cooling. 
	The emergence of new peaks in neutron diffraction below $T_2=54.5$\,K are consistent with magnetic ordering. 
	
	
	Additional magnetic Bragg peaks due to magnetic ordering of the 2.8(3)wt\% Mn$_3$O$_4$ impurity at $\sim$43\,K~\cite{boucher1971magnetic,jensen1974magnetic} are observed. 
	At 40\,K and 2.5\,K, peaks at $\sim$28.4$^\circ$ ($d=4.9$\,\AA{}), $\sim$34.7$^\circ$ ($d=4.04$\,\AA{}; little more than a shoulder at 40\,K), and $\sim$46$^\circ$ ($d=3.08$\,\AA{}) are attributed to the [101], [110], and [112] magnetic reflections respectively~\cite{boucher1971magnetic,boucher1971proprietes,jensen1974magnetic,kemei2014structural}. 
	At 2.5\,K, Mn$_3$O$_4$ has extra Bragg peaks due to its magnetic transition at 33\,K~\cite{boucher1971magnetic,boucher1971proprietes,jensen1974magnetic,kemei2014structural}.
	
	We assume all other emergent peaks are attributable to the cryptomelane phase. 
	The following two sections will detail our analysis of the $T_3<T<T_2$ regime and the $T<T_3$ magnetic regime. 
	Due to the low resolution of the neutron data compared with that of the synchrotron, the subtle monoclinic distortion cannot be resolved in this data. 
	We still perform magnetic analysis using the monoclinic structure, but present alternative analysis using the tetragonal structure in SI Section~9. 
	
	\subsection{Magnetic structure for $T_3<T< T_2$}

	At 47\,K and 44\,K, five clear magnetic Bragg peaks are observed in the difference pattern [Figure~\ref{diff_patterns}].
	Only one of the main magnetic peaks, at $d=6.95$\,\AA, can be assigned to a commensurate magnetic structure, with propagation vector $\vec{k}_\mathrm{mag}=(0,0,0)$. 
	This peak corresponds to the nuclear [110] Bragg peak in the tetragonal basis, or a superposition of the [100] and [$\bar{1}$20] reflections in the $A2/m$ monoclinic unit cell. 
    We can be certain this is magnetic in origin because (1) there is no increase in nuclear intensity at higher-2$\theta$ as we would expect if it were structural in origin, and (2) it exhibits intensity increases on cooling expected of a magnetic peak [Figure~S5]. 
    We also see a very subtle feature on the low-$d$ side of the incommensurate $d=5.15$\,\AA{} peak which becomes slightly more prominent in the data at 44\,K than at 47\,K, and also gets fit by the commensurate component of the magnetic structure. 
	The remaining four peaks ($d=5.15,4.10,3.15,2.65$\,\AA) can only be indexed to a propagation vector incommensurate with the crystal structure, $\vec{k}_\mathrm{mag}=(0,0,k_z)$ [$k_z\approx0.3690(2)$] (where we here use a monoclinic cell with special $c$-axis).

\begin{figure}[t]
		\includegraphics[width=87mm]{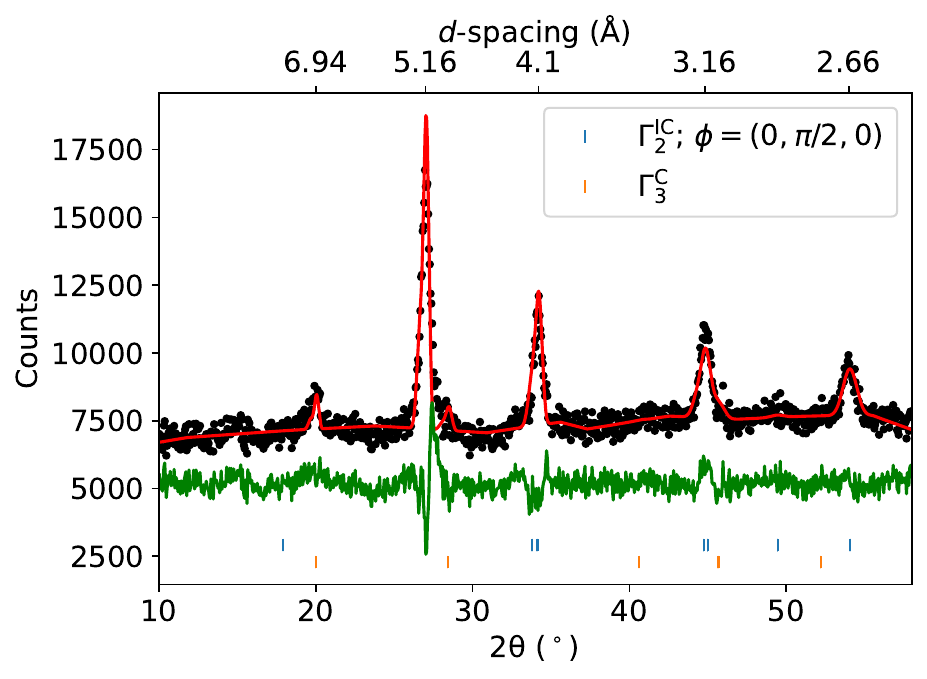}
		\caption{
			\label{fig_mag_refinement}
			Experimental difference pattern 47\,K-60\,K (KMO-S1), compared with calculated pattern obtained by Rietveld refinement of the model for $T_3<T<T_2$, consisting of a commensurate $|\vec{k}_\mathrm{mag}|=0$ and incommensurate $\vec{k}_\mathrm{mag}=(0,0,\sim 0.37)$ component, 
			assuming a monoclinic ($A2/m$) unit cell. 
			The incommensurate structure was represented with a helical model, with moments only in the plane perpendicular to the propagation vector. 
			We note that a fit of comparable quality can be obtained by substituting the incommensurate component with either a sinusoidal modulation in monoclinic symmetry [SI Figure~S29], an ellipsoidal helix [Figure~S31], or a constant-moment helix assuming tetragonal ($I4/m$) symmetry [SI Figure~S34]. 
		}
	\end{figure}

	For each propagation vector, there are multiple symmetry-allowed IRs. The correct IR would describe the basis vector behaviors at each magnetic site, which collectively make up the magnetic structure. 
	There are 2 possible IRs for $\vec{k}_\mathrm{mag}=(0,0,\sim0.37)$ and 4 IRs for $\vec{k}_\mathrm{mag}=(0,0,0)$ in a $A2/m$ cell with magnetic ions on the $4i$ Wyckoff site; these are denoted using $\Gamma_i^j$ ($i=1,2,3,$ or $4$; $j=\mathrm{IC}$ or $\mathrm{C}$ where IC/C stands for InCommensurate or Commensurate). 
    We compare the properties of these in Table~\ref{tab:irreps}. 
	The basis vectors (BVs) are tabulated in Section 7 of Supplementary Information. 
	We then use magnetic Rietveld refinement to test the IRs against the experimental difference pattern (47\,K-60\,K), shown in Figure~\ref{fig_mag_refinement}, and determine the magnetic structure. 
	
	Of the four IRs for the commensurate component, $\Gamma_1^\mathrm{C}$ and $\Gamma_3^\mathrm{C}$ give a net moment, and $\Gamma_2^\mathrm{C}$ and $\Gamma_4^\mathrm{C}$ do not (see Table~S6 of SI). 
	As we know from magnetometry that there is a net moment in the $T_3<T<T_2$ regime, and this net moment must come from the commensurate component of the magnetic structure, we only consider $\Gamma_1^\mathrm{C}$ and $\Gamma_3^\mathrm{C}$ as possible IRs. 
	We then consider prior experimental work on K$_{1.5}$(H$_3$O)$_x$Mn$_8$O$_{16}$ single crystals which indicate that the magnetic hysteresis is much stronger in the plane perpendicular to the tunnels than along the tunnels~\cite{sato1999charge}. 
	On this basis, we then rule out $\Gamma_1^\mathrm{C}$ which only has moments parallel to the nanotunnels, leaving only $\Gamma_3^\mathrm{C}$ as a viable candidate for the commensurate component of the magnetic structure. 
	In this IR, all moments at a given site are collinear. 
	From preliminary analysis, there appears no improvement in the fit from decoupling the moments on the two Mn sites in this $|\vec{k}_\mathrm{mag}|=0$ structure, so we constrain the moments on both Mn sites to be equal in magnitude. 
	We find that: (1) fit quality is entirely independent of the direction of the net moment within the plane perpendicular to the tunnels, and (2) refining the basis vectors in the [100] and [010] directions independently leads to divergence within \textsc{FullProf}. 
	This is due to the fact that, to the low resolution of the neutron diffraction data, the monoclinic splitting cannot be resolved and therefore our cell is pseudo-tetragonal; in a tetragonal cell, the direction of moments in the basal plane cannot readily be determined using powder diffraction data. 
	For this reason, we arbitrarily constrain the net moments in the $a$ and $b$-directions to be equal. We do not assert that net moment must occur in the [110] direction for the commensurate component because a [100] or [010] net moment would also be consistent with the available data, and for this reason we do not show a figure of the commensurate structure. 
	For completeness, we note that the analysis using a tetragonal cell (Section~S9) yields a similar commensurate component to that found with the monoclinic cell. 

    \begin{table}[ht]
        \centering
        \caption{
            Symmetry-allowed irreducible representations (IRs) for the monoclinic ($A2/m$) cell and their consistency with experimental constraints (see also Tables~S5 and S6 of SI). 
            The nanotunnel direction is $\parallel c$; the perpendicular plane is the $ab$-plane. 
            ``FM" and ``AFM" under ``Net moment" indicate ferromagnetic and antiferromagnetic moment symmetry respectively. 
            Under ``diffraction fit quality", only IRs which cannot be ruled out on the basis of moment direction or net moment are tested. 
            For the incommensurate IRs, we assume moments are in the $ab$-plane to enforce helical ordering. 
            Selected IRs are shown in bold. 
        }
        \label{tab:irreps}
        \begin{tabular}{ll|lcc}
            \hline\hline
            $\vec{k}_\mathrm{mag}$ & IR & Moment direction & Net moment & Fit quality \\
            \hline
            \multirow{4}{*}{$(0,0,0)$}
              & $\Gamma_1^\mathrm{C}$ & $\parallel c$ & FM & \textbf{--} \\
              & $\Gamma_2^\mathrm{C}$ & $\parallel ab$-plane & AFM & \textbf{--} \\
              & \textbf{$\Gamma_3^\mathrm{C}$} & $\parallel ab$-plane & FM & Good \\
              & $\Gamma_4^\mathrm{C}$ & $\parallel c$ & AFM & \textbf{--} \\
            \hline
            \multirow{2}{*}{$(0,0,{\sim}0.37)$}
              & $\Gamma_1^\mathrm{IC}$ & $\parallel ab$-plane & AFM & Poor \\
              & \textbf{$\Gamma_2^\mathrm{IC}$} & $\parallel ab$-plane & AFM & Good \\
            \hline\hline
        \end{tabular}
    \end{table}

\begin{figure*}[hbtp]
		\includegraphics[width=\linewidth]{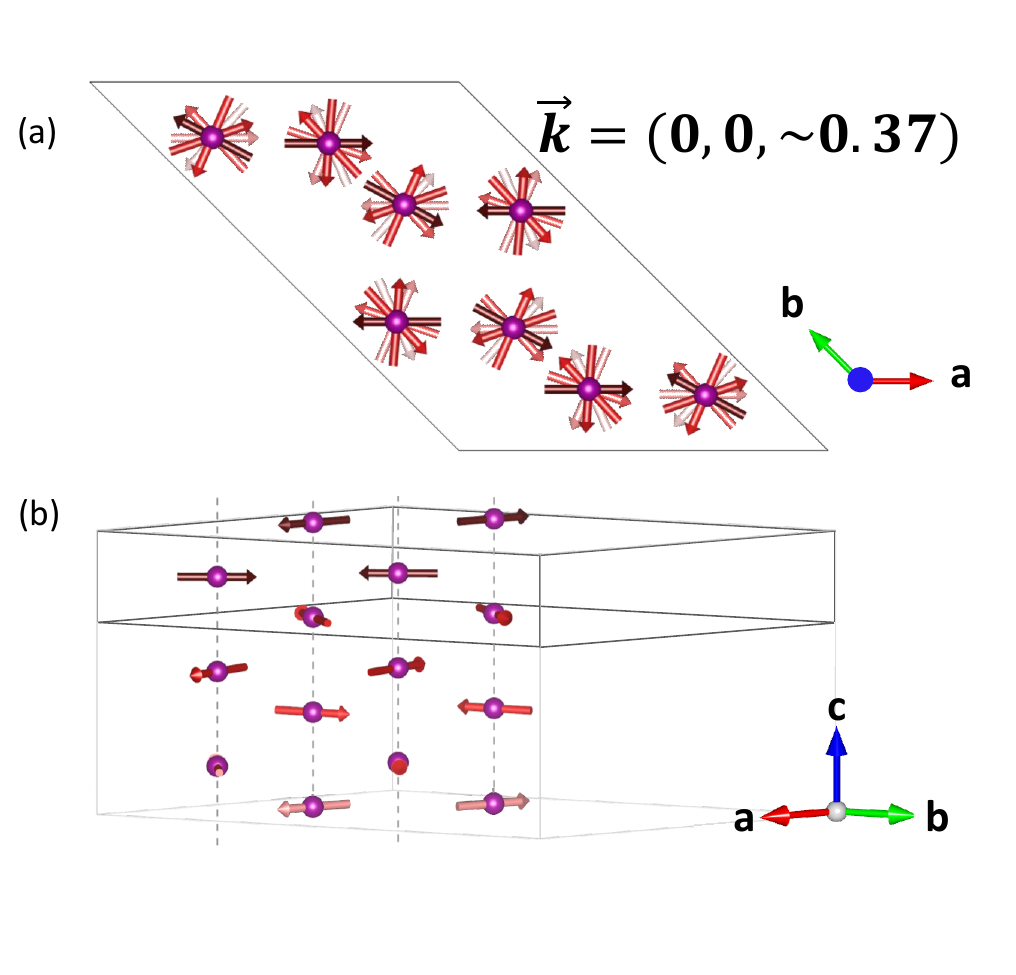}
		\caption{
			\label{monoclinic-mag-structure.pdf}
			The refined magnetic unit cell (KMO-S1) for the incommensurate $\vec{k}_\mathrm{mag}=(0,0,\sim0.37)$ components of the magnetic structure for $T_3<T<T_2$ assuming a monoclinic ($A2/m$) unit cell with helical ordering. 
			(a) View perpendicular to nanotunnels, (b) view along nanotunnels. 
		}
	\end{figure*}

	For the incommensurate $\vec{k}_\mathrm{mag}=(0,0,\sim0.37)$, it is possible to obtain a reasonable fit to the incommensurate peaks in the experimental diffraction data with $\Gamma_2^\mathrm{IC}$, but not with $\Gamma_1^\mathrm{IC}$. 
	We first tested a $\Gamma_2^\mathrm{IC}$ model assuming sinusoidal modulation of moments along the nanotunnels, which yields a good fit [SI Figure~S29].
	However, we noted that there appeared no dependence whatsoever of fit quality on the plane in which moments are allowed to oscillate sinusoidally; as with the invariance of fit quality with moment direction for the commensurate case, this is a consequence of symmetry. 
	This observation, combined with the prior experimental and theoretical works which indicate helical behavior in this temperature regime~\cite{mandal2014incommensurate,sato1997magnetism,sato1999charge}, along with the implausibility of Mn cations with zero moment, led us to test a helical model in which constant moments rotate about the propagation vector with period equal to $c/|\vec{k_\mathrm{mag}}|$. 
	This was achieved by representing moments as spherical coordinates, and configuring a 180$^\circ$ phase shift between the two Mn sites to ensure that the structure was equivalent to $\Gamma_2^\mathrm{IC}$. 
	The resulting fit is shown in Figure~\ref{fig_mag_refinement} and the associated magnetic structure is shown in Figure~\ref{monoclinic-mag-structure.pdf}(b). 

	While the coexistence of two possible IRs may indicate magnetic phase segregation into two phases each with one of these two IRs, we do not consider this as likely in this context. 
	Firstly, given that the sample has multivalent Mn, magnetic phase segregation would probably correspond to segregation of Mn$^{3+}$ and Mn$^{4+}$ which would be identifiable as phase segregation in high-resolution synchrotron diffraction. 
	Secondly, if magnetic phase segregation were to occur, this would likely be a consequence of chemical inhomogeneity, and therefore synthesis-dependent, and yet we see very similar magnetic properties for $T_3<T(\mathrm{K})<T_2$ reported in cryptomelane prepared via different synthesis routes, including single crystal synthesis~\cite{sato1997magnetism} and hydrothermal synthesis~\cite{tseng2015magnetic}. 
	Additionally, we note that coexistence of a $|\vec{k}_\mathrm{mag}|=0$ and $|\vec{k}_\mathrm{mag}|\neq0$ component is not uncommon in magnetic structures~\cite{ivanov2012temperature,cooley2020evolution}. 
	We therefore consider it likely that both IRs coexist over the bulk of the sample, and hence fit the scale factors $S_\mathrm{C}=S_\mathrm{IC}=S_\mathrm{crystal}$ from which the refinement determines the magnitude of each moment for the commensurate component and the magnitude of the individual moments in the incommensurate component (assuming purely helical modulation). 
	The net magnitude of each individual moment would result from the superposition of the commensurate and incommensurate components; at 47\,K, the magnitude for each moment occurs in the range 0.513(4)$\pm$0.320(16)\,$\mu_\mathrm{B}$ / Mn depending where along the modulation the moment occurs, increasing to 0.701(4)$\pm$0.496(20)\,$\mu_\mathrm{B}$ / Mn on cooling to 40\,K [Table~S9]. 
    Theoretically, magnetic moments obtained by Rietveld refinement should converge at 0\,K with $\mu_\mathrm{sat}=gS$, which is typically smaller than the $\mu_\mathrm{eff}$ obtained by Curie-Weiss analysis, and for Mn$^{3+}$ or Mn$^{4+}$ gives 4\,$\mu_\mathrm{B}$ / Mn and 3\,$\mu_\mathrm{B}$ / Mn respectively, or $\sim$3.2\,$\mu_\mathrm{B}$ / Mn for our stoichiometry. Clearly the value we obtain at 47\,K is substantially smaller than this. 
    Figure S5 shows that the peak intensities significantly increase on cooling to lower temperatures, which is consistent with the rapid increase in magnetization seen from Figure~\ref{fig-mag}(a) on cooling from 47\,K to the temperature just above $T_3$ (i.e. around 28\,K).
    
    The suppression of the magnetic moment determined from neutron diffraction is compounded by several additional factors well-documented in frustrated Mn oxide systems: covalency between Mn \textit{d}-orbitals and O \textit{p}-orbitals~\cite{streltsov2016covalent} and geometric frustration~\cite{larson2015inducing} and hence at 47\,K a moment of $\sim$0.5\,$\mu_\mathrm{B}$ / Mn is not unreasonable. 
    Similar behavior is observed in Ba$_{1.2}$Mn$_8$O$_{16}$ where refinement of the magnetic structure at 5\,K gives a moment of 2.3\,$\mu_\mathrm{B}$ / Mn, which is smaller than the $\mu_\mathrm{eff}\approx$4.34\,$\mu_\mathrm{B}$ / Mn determined from Curie-Weiss analysis of the magnetic susceptibility~\cite{larson2015inducing}. 
    
	The obtained moment from the commensurate component appears approximately consistent with that obtained from the hysteresis loops from the fixed-temperature, variable-field magnetometry measurement [Figure~\ref{MvH-fig}]. 
							
	We note that we observe no dependence of the propagation vector on temperature in the incommensurate magnetic component between 40\,K and 47\,K [Figure~S6], and so it is unlikely there are any qualitative changes in the magnetic structure within this temperature regime. 

	\subsection{Magnetic structure for $T<T_3$}

	For $T<T_3$ we only have one diffraction pattern, at 2.5\,K. 
	There are many more peaks in the 2.5\,K-60\,K difference pattern than in the 47\,K-60\,K difference pattern, although the peaks due to the $\vec{k}_\mathrm{mag}=(0,0,\sim0.37)$ component disappear entirely (see Figure~\ref{diff_patterns}). 
	This increase in the number of peaks is in part due to the magnetic ordering of Mn$_3$O$_4$, which exhibits several additional magnetic peaks due to its commensurate magnetic structure below $\sim33$\,K~\cite{boucher1971magnetic,boucher1971proprietes,jensen1974magnetic,kemei2014structural}. 
	Despite this, most of the peaks are not attributable to a known impurity and are therefore attributed to a distinct magnetic ordering in cryptomelane. 
	
	It should be noted here that it has been reported that at low temperatures, the impurity phase Mn$_3$O$_4$ exhibits some structural phase separation, with a fraction of it exhibiting an orthorhombic structure as opposed to its usual tetragonal structure~\cite{kim2011pressure,chung2013low,kemei2014structural}. 
	However, such effects are too subtle to observe when Mn$_3$O$_4$ is a minor impurity, and we do not include the reported orthorhombic phase in our analysis.

	Similarly to $T_3<T<T_2$, the majority of the magnetic peaks for $T<T_3$ cannot be indexed to a commensurate magnetic unit cell. 
	However, the nuclear [110] and [220] Bragg peaks do increase in intensity at 2.5\,K compared with 60\,K, and so it is likely that, as is the case for $T_3<T<T_2$, there is a commensurate component to the magnetic ground state. 
	Unlike the higher-temperature commensurate component, such a component cannot have a net moment, as inferred from the lack of hysteresis for $T<T_3$ from the fixed-temperature variable-field magnetization measurements [Figure~\ref{MvH-fig}]. 
	The commensurate component is therefore likely to be described by either $\Gamma_2^\mathrm{C}$ and $\Gamma_4^\mathrm{C}$ [Table~S6]. 

	Excluding these commensurate peaks, we have attempted to index the remaining peaks. 
	However, we have not been able to find a single incommensurate propagation vector with fewer than 3 incommensurate components which leads to all additional peaks being fit. 
	We note that $\vec{k}_\mathrm{mag}=(0,0,\sim0.15)$ is able to fit some of the additional peaks, but not all. 
	This means that either the propagation vector is incommensurate in 3 spatial dimensions, or that there are multiple incommensurate propagation vectors. 
	Both of these possibilities are difficult to test with neutron powder diffraction alone, and single-crystal studies are required. 
	Consequently, we are not able to solve the magnetic structure for $T<T_3$, but we can say that it is more complex than for $T_3<T<T_2$ and that a purely canonical spin glass cannot be the only bulk state. 


	\section{\label{sec:level5}Discussion}

	To understand the magnetic properties of a material, it is important to understand the nature of the exchange interactions.
	For $J_1$ and $J_2$ interactions it is unlikely there is direct exchange given the large interatomic distances relative to the Shannon radii~\cite{shannon1970revised} of Mn$^{3+}$ and Mn$^{4+}$. 
	Given the presence of both Mn$^{3+}$ and Mn$^{4+}$, rationalizing the nature of the interactions using the Goodenough-Kanamori-Anderson rules~\cite{goodenough1955theory,goodenough1958interpretation,kanamori1959superexchange} is non-trivial~\cite{crespo2013competing}. 
	This is further complicated by the possibility for double exchange; while the Mn-O-Mn angles deviate from 90$^\circ$, double exchange has been discussed in other mixed valent materials with non-90$^\circ$ \textit{M}-O-\textit{M} (\textit{M} = transition metal) bond angles, such as rutile chains~\cite{dutton2010divergent}.

	Our finding that the magnetic ordering length-scale is incommensurate with the lattice parameters in K$_{1.448(3)}$Mn$_8$O$_{16}$ is in keeping with prior literature for doped- and undoped-$\alpha$-MnO$_2$. 
	For example, a theoretical study on the undoped variant using a Heisenberg model~\cite{mandal2014incommensurate} does indeed predict helical ordering. 
	The prediction in the study~\cite{mandal2014incommensurate} requires the angle of helical rotation of spins per lattice parameter, $\phi$, to meet the condition $\pi/2 <\phi <\pi$, which in our case it does, as $\phi\approx0.37\cdot2\pi=0.74\pi$. 
	This places the observed incommensurate modulation within the helical regime predicted for frustrated interactions on the hollandite lattice, supporting a qualitative correspondence between the theoretical model and the experimentally observed modulated magnetic order for $T_3<T<T_2$. 

    As mentioned, we also observe an incommensurate structural distortion with $\vec{k}_\mathrm{struc}=0.740834(7)\vec{c^*}$. 
    Similar distortions have been observed for other hollandites~\cite{Cheary1992Electron16,carter2005universally,larson2017metal} and are typically attributed to cation ordering (in this case K$^+$) within the nanotunnels, though some cryptomelane-focused works which do not include diffraction have attributed the associated transition with Mn$^{3+/4+}$ charge ordering~\cite{sato1999charge,luo2010tuning}. 
    Carter and Withers (2005)~\cite{carter2005universally} proposed\footnote{We note they actually used $\vec{b^*}$ due to a different crystallographic setting in their work.} for barium titanium hollandites that $2\vec{k}_\mathrm{struc}=x\vec{c^*}$ for $x$ in $A_x$Ti$_8$O$_{16}$. 
    Using this equation for cryptomelane, we can back-calculate stoichiometry from our determined structural propagation vector, we obtain a value ($x=1.48167(1)$) close to the composition obtained based on impurity fractions for KMO-S2 ($x=1.488(2)$). 
    This numerical agreement is consistent with Carter and Withers' equation~\cite{carter2005universally} derived for a chemically distinct barium titanium hollandite, supportive of a common origin for the incommensurate structural order in cryptomelane and in the barium titanium hollandites (i.e. cation ordering~\cite{Cheary1992Electron16,carter2005universally}). 
    
	There is also a conceptual correspondence between our work and the helical ferrimagnetism model proposed by Sato \textit{et al.}~\cite{sato1997magnetism,sato1999charge} for K$_{1.5}$(H$_3$O)$_x$Mn$_8$O$_{16}$. 
	In this model, there are modulated spin and charge behaviors with length-scales $\lambda_\mathrm{m}$ and $\lambda_\mathrm{c}$ respectively, along the crystallographic axis of the tunnels. 
	The modulated charge behavior is due to mixed valent Mn$^{3+}$/Mn$^{4+}$. 
	The charge modulation will determine the magnitude, and the spin modulation will determine the orientation, of individual spins. 
	For a normal helical magnetic system, there will be no net moment, but when $\lambda_\mathrm{c}/\lambda_\mathrm{m}=n$ ($n$ being any integer), with this combination of charge and spin modulations it is argued that there will be a slight net magnetic moment. 
	Sato \textit{et al.}~\cite{sato1997magnetism,sato1999charge} hypothesize that for $25<T($K$)<55$, $n$ is an integer, and the transition at $T\approx25$\,K results in $n$ no longer having an integer value, hence the lack of a net moment below this temperature. 
	This model, for $T_3<T<T_2$, would be best represented in terms of propagation vectors by a dual-$\vec{k}$ model with $\vec{k}_\mathrm{mag}=(0,0,k_z)$ and $\vec{k}_\mathrm{mag}=(0,0,0)$, which is what we observe in KMO-S1. 
    However, in their work they suggest that $\vec{k}_\mathrm{struc}=0.125$ implicitly, whereas we have an incommensurate propagation vector. 
    To ensure we are comparing vectors for the same sample, we calculate $\vec{k}_\mathrm{struc}=0.724(3)$ for KMO-S1 using the Rietveld-refined composition and the relation proposed by Carter and Withers~\cite{carter2005universally}. 
    
    This leads to $ \lambda_\mathrm{c}/\lambda_\mathrm{m} = |\vec{k}_\mathrm{mag}/\vec{k}_\mathrm{struc}| = 0.3690/0.724=0.5097$ which is not an integer, though it may be worth noting that it is very close to a ratio of integers 1:2. 
    We therefore conclude that our results are conceptually similar to the helical ferrimagnetism model of Sato \textit{et al}~\cite{sato1997magnetism,sato1999charge} in terms of the magnetic components we use, but that the mechanistic origin cannot match the model because the ratio between propagation vectors does not follow that logic. 

	Crespo \textit{et al.}~\cite{crespo2013competing} show theoretically that the $\alpha$-MnO$_2$ crystallographic motif with inhomogeneous doping may exhibit spin glass behavior in the presence of antiferromagnetic exchange interactions in an Ising spin lattice. 
	Previous experimental works on K-rich cryptomelane have indicated that $T_3$ may be a spin glass-like transition~\cite{barudvzija2016structural,barudvzija2020magnetic}. 
	From our DC and AC susceptibility measurements, we see an FC/ZFC divergence which could support this.
	However, we do not see a strong frequency dependence to the magnetic behavior [Figure~\ref{AC-fig}] which is a characteristic feature of a spin glass~\cite{cannella1972magnetic,mulder1981susceptibility,binder1986spin,shirakami1998spin,khan2024spin,freedberg2024brief}; though we note that the $T_3$ feature is sufficiently weak in our AC susceptibility data that a quantitative frequency-shift or relaxation-time analysis of the kind used to characterize canonical spin glasses is not supported by these data. 
	Furthermore, the magnetic Bragg peak structure we see for $T<T_3$ is not consistent with canonical spin glass-like ordering, as a spin glass phase would not exhibit magnetic Bragg peaks~\cite{binder1986spin}. 
	The proposed transition to a canonical spin glass-like state is therefore not supported by our data. 
    While a re-entrant spin glass superposing glassiness over magnetic order may be consistent with the Bragg peaks we observe, we do not see any evidence for this. 
	The possibility that spin glass-like behavior occurs at the surface has been suggested for similar materials, including Na$_{2-x}$Mn$_8$O$_{16}$~\cite{lan2011synthesis} and Ba$_{1+\delta}$Mn$_8$O$_{16}$~\cite{yu2010spin}, and cannot be ruled out based on our data. 
	The possibility that a spin glass is caused by variations in local K$^{+}$ content within some regions of the sample is also highly plausible; it is already known that spin glass-like behavior occurs for K$_x$Mn$_8$O$_{16}$ ($x<1$)~\cite{suib1994magnetic,luo2009spin,luo2010tuning}, and it has also been shown that Sn-doping of $\alpha$-MnO$_2$ results in spin glass-like behavior~\cite{liu2022tuning}; such behavior may speculatively resemble a cluster spin glass, though we do not have any evidence for this. 
	Therefore, we suggest that our observed magnetic Bragg peaks for $T<T_3$ may be reconcilable with previous claims of spin glassiness, but that such a spin glass-like state cannot be the bulk ground state of KMO-S1.

	Our findings of magnetic Bragg peaks in both temperature regimes is consistent with the only known previous neutron diffraction study on K$_{1.72}$Mn$_8$O$_{16}$~\cite{larson2017frustrated_chapter}. 
	However, the magnetic peak positions do not match those in this study, with the result that our magnetic structure solution is unlikely to be the magnetic structure solution for their reported data, although there is the possibility for a $\vec{k}_\mathrm{mag}=(0,0,0)$ component from their data, as for ours. 
	This indicates that there is a strong dependence of the magnetic propagation vector(s) on Mn:K ratio, similar to the structural propagation vector, which should be studied further in future work.

	The first magnetic ordering temperature of cryptomelane, at around 55\,K, is around 13\% the value of the Weiss temperature of $-425$\,K reported by Larson~\cite{larson2017frustrated_chapter} for the closely-related composition K$_{1.33}$Mn$_8$O$_{16}$; we observe something similar in our own Curie-Weiss analysis (SI Section~11). 
	This means the magnetic ordering occurs at a far lower energy scale than that of the magnetic interactions, suggesting that the ordering is inhibited by the frustrated magnetic interactions present in the Mn sublattice of the crystal structure. 
	From this we can conclude that frustration likely plays a role in the magnetism, an inference further supported by the spin reorientation around 25\,K. 
	There are parallels with this in the literature; for instance, MgCr$_2$O$_4$~\cite{bai2019magnetic} and LiInCr$_4$O$_8$~\cite{nilsen2015complex} spinels, and CaMn$_7$O$_{12}$~\cite{przenioslo1999magnetic}, are frustrated systems exhibiting two co-existing propagation vectors. 
	The presence of multiple propagation vectors is likely a consequence of these competing interactions with large energy scales. 

	\section{\label{sec:level6}Conclusion}

	We have studied the structural and magnetic properties of cryptomelane, K$_x$Mn$_8$O$_{16}$ (KMO-S1) with precise composition $x=1.448(3)$. 
	We have found that, structurally, the previously reported transition $T_1$ corresponds to a structural transition: below this temperature, there emerges a subtle monoclinic distortion from the high-temperature tetragonal aristotype, and an incommensurate structural ordering along the nanotunnels; in a sample with a slightly higher K content we measured $\vec{k}_\mathrm{struc}=0.740834(7)\vec{c^*}$. 
	Examining the magnetic properties, we find that there are two distinct ordered magnetic regimes in the temperature ranges $T<24$\,K and $24<T\mathrm{(K)}<54.5$. 
	For $T<24$\,K, the sample has long-range magnetic ordering, so a canonical spin glass cannot be the bulk ground state in this temperature regime. 
	For $24<T<54.5$\,K, there is an incommensurate propagation vector $\vec{k}_\mathrm{mag}=(0,0,k_z)$ [$k_z=0.3690(2)$], which is likely helical in the $ab$-plane, along with a commensurate magnetic cell which we successfully fit with a $\vec{k}_\mathrm{mag}=(0,0,0)$ propagation vector. 
	This is consistent with the predictions of Mandal \textit{et al}~\cite{mandal2014incommensurate}. 
    The result is similar to the predicted ``helical ferrimagnetism" of Sato \textit{et al}~\cite{sato1997magnetism,sato1999charge}, though the ratio between $\vec{k}_\mathrm{mag}$ and $\vec{k}_\mathrm{struc}$ doesn't quite match their predictions. 

	Future work would consist of single-crystal diffraction studies on cryptomelane. 
	Single-crystal neutron diffraction would be used to identify the nature of the magnetic ordering below $T_3\approx24$\,K, and to attempt to distinguish between helical, sinusoidal, and conical magnetic modulations for the incommensurate component of the magnetic structure for $T_3<T<T_2$. 
	Additionally, neutron studies for a range of K$^{+}$ stoichiometries would be needed to determine the compositional dependence of the magnetic structure.




	\section*{Funding}
	
	LNC and NDK acknowledge a scholarship EP/R513180/1 to pursue doctoral research from the UK Engineering and Physical Sciences Research Council (EPSRC). 
	LNC also received funding from the Cambridge Philosophical Society and the Faraday Institution (FIRG060).
	KS acknowledges funding from the British Council's Newton-Bhabha Fund PhD Placements Programme, which enabled him to visit the University of Cambridge on a research placement. 
	JDB acknowledges support from the Robert A. Welch Foundation, grant (E-2179-20240404). 
	PB is grateful to Alexander von Humboldt Foundation (Bonn, Germany) for a 2022 Humboldt fellowship for experienced researchers. 
	SED acknowledges funding from EPSRC (EP/T0285580/1). 
	
	\section*{Acknowledgments}

	Magnetic and heat capacity measurements at the University of Cambridge were made on the EPSRC Advanced Characterization Suite (funded under EP/M0005/24/1).
	We acknowledge Diamond Light Source for time on the I11 instrument under BAG proposal CY28349 (December 2021) and rapid access proposal CY33664-1 (October 2022).
	We acknowledge the Institut Laue-Langevin (ILL) for neutron diffraction measurements on the high-intensity powder diffractometer D20 and the high-resolution powder diffractometer D2B under proposal EASY-893.

	LNC thanks Dr Saptarshi Mandal and Dr Efrain E. Rodriguez for useful correspondence.

	\section*{Data availability}
	The variable-temperature neutron diffraction data from the ILL is available at doi:10.5291/ILL-DATA.EASY-893~\cite{doi:10.5291/ILL-DATA.EASY-893}. 
	The remaining data is available in the University of Cambridge repository at doi:10.17863/CAM.134500~\cite{cam_data}.

	\section*{Supplementary Information}
	A document containing Supplementary Information to this article is available. 
    Additional experimental details, sample characterization, diffraction/magnetometry analysis, electron microscopy images, low-temperature neutron diffraction, laboratory X-ray, and synchrotron X-ray diffraction data and refinements, magnetic structure models and irreducible representations, heat capacity and Curie-Weiss analysis methodology.

	\bibliography{references}

\end{document}